\documentclass[amssymb,twocolumn,prb,notitlepage,superscriptaddress,floats,amsmath]{revtex4-1}

\usepackage{epsfig}
\usepackage{bm}
\usepackage{color}
\usepackage{array}

\begin{document}

\title{Semiconductor Bloch equation analysis of optical Stark and Bloch-Siegert shifts \\
in monolayers WSe$_2$ and MoS$_2$}
			
\author{A.O. Slobodeniuk}
\email{aslobodeniuk@karlov.mff.cuni.cz}
\affiliation{Department of Condensed Matter Physics, Faculty of Mathematics and Physics,
Charles University, Ke Karlovu 5, CZ-121 16 Prague, Czech Republic}

\author{P. Koutensk\'{y}}
\affiliation{Department of Chemical Physics and Optics, Faculty of Mathematics and Physics,
Charles University, Ke Karlovu 3, CZ-121 16 Prague, Czech Republic}

\author{M. Barto\v{s}}
\affiliation{Central European Institute of Technology, Brno University of Technology, Purky\v{n}ova 656/123, 612 00 Brno, Czech Republic}

\author{F. Troj\'{a}nek}
\affiliation{Department of Chemical Physics and Optics, Faculty of Mathematics and Physics,
Charles University, Ke Karlovu 3, CZ-121 16 Prague, Czech Republic}

\author{P. Mal\'{y}}
\affiliation{Department of Chemical Physics and Optics, Faculty of Mathematics and Physics,
Charles University, Ke Karlovu 3, CZ-121 16 Prague, Czech Republic}

\author{T. Novotn\'{y}}
\affiliation{Department of Condensed Matter Physics, Faculty of Mathematics and Physics,
Charles University, Ke Karlovu 5, CZ-121 16 Prague, Czech Republic}

\author{M. Koz\'{a}k}
\email{kozak@karlov.mff.cuni.cz}
\affiliation{Department of Chemical Physics and Optics, Faculty of Mathematics and Physics,
Charles University, Ke Karlovu 3, CZ-121 16 Prague, Czech Republic}


\begin{abstract}
We report on the theoretical and experimental investigation of valley-selective optical Stark and Bloch-Siegert shifts of exciton resonances in monolayers WSe$_2$ and MoS$_2$ induced by strong circularly polarized nonresonant optical fields. We predict and observe transient shifts of both 1sA and 1sB exciton transitions in the linear interaction regime. The theoretical description is based on semiconductor Bloch equations. The solutions of the equations are obtained with a modified perturbation technique, which takes into account many-body Coulomb interaction effects. These solutions allow to explain the polarization dependence of the shifts and calculate their values analytically. We found experimentally the limits of the applicability of the theoretical description by observing the transient exciton spectra change at high field amplitudes of the driving wave.
\end{abstract}


\maketitle

\section{Introduction}

Energy band structure of a crystalline solid may contain multiple energy degenerate minima of the band gap. The electrons localized in the individual minima possess a valley degree of freedom in addition to charge and spin. The energy valleys are usually separated by a large crystal momentum leading to relatively long intervalley scattering times \cite{Hammersberg2014} potentially allowing to utilize the valley quantum number for information processing and storage.

Generation, manipulation and readout of unbalanced valley populations are possible via several processes. The first mechanism is based on selective optical excitation and can be applied in materials, in which the optical selection rules connect the excitation of carriers in different valleys to a certain polarization state of light. An example of such materials are two-dimensional transition metal dichalcogenide (2D TMDs) monolayers \cite{Mak2010,Splendiani2010}, where valley-selective optical excitation of excitons \cite{He2014,Chernikov2014} in K$^+$ or K$^-$ points of the Brillouin zone can be reached using circularly polarized resonant light \cite{Xiao2012,Cao2012,Yao2008}.

Another mechanism exploits the anisotropy of effective masses of carriers in different groups of valleys. The excited electrons and holes accelerated by static electric field can reach kinetic energy, which is required for intervalley scattering mediated by the interaction with phonons. The carriers with low effective mass in the direction of the applied field gain higher kinetic energy than heavier quasiparticles and therefore the probability of intervalley scattering from light to heavy valleys is larger than in the opposite direction. This mechanism allows to generate valley polarized electron population from the initial isotropic distribution in momentum space, in diamond 
\cite{Isberg2013,Suntornwipat2021}. 

The third mechanism reaches valley-selective control by lifting the energy degeneracy between different groups of valleys using static electric or magnetic fields or coherent optical phenomena such as the optical Stark (OS) 
\cite{Autler1955,Ritus1967,Schuda1974,Ell1989,Lindberg1988,Chemla1989,VonLehmen1986,Delone1999} or 
the Bloch-Siegert (BS) shifts \cite{Bloch1940,Stevenson1940,Shirley1965,Allen1975}. 

The control of valley degrees of freedom with static magnetic and electric fields has been clearly demonstrated in TMD monolayers \cite{Wang2016,Aivazian2015,Molas2019_1}. However, it turned out that even strong fields can produce relatively small energy shifts, e.g, 1-2~meV for magnetic fields at 30~T \cite{Molas2019_1,Stier2018,Goryca2019}. 
Additional conditions are usually required to observe such tiny shifts such as helium temperatures,
the unique sources of static fields, and the state-of-art setups/detectors. Moreover, the static fields can't provide 
the real-time dynamical control of valley degrees of freedom in solids. 
Therefore, despite the static fields can demonstrate the possibility of the manipulation of the valley degrees of freedom in crystals, they are not suitable for realistic applications.      
This problem can be solved with the help of time-varying electromagnetic fields, e.g., with light beams. 

Previously it was demonstrated that light pulses can provide coherent control of various electronic systems, offering high-speed and nondestructive mechanisms for quantum measurement and manipulation 
\cite{Gupta2001,Press2008,Berezovsky2008}. The key ingredient of such mechanisms is the application of nonresonant light, which induces OS and/or BS shifts of energy levels in the system without exciting real population.  
By these means, the OS and BS effects can be applied to control the energy levels in atoms and molecules \cite{Autler1955,BonchBruevich1968}, but may also be used in solid state systems consisting of quantum wells, dots or bulk semiconductor materials \cite{Mysyrowicz1986,VonLehmen1986,Joffre1988}, and finally in relatively recently discovered
2D semiconductors, like TMD monolayers \cite{Kim2014,Sie2015,Sie2017,Cunningham2019,LaMountain2018}. 

For small detuning $\delta=|E_0-\hbar\omega|\ll E_0$ between the photon energy of the non-resonant driving light 
$\hbar\omega$ and the energy of the resonance $E_0$, 
the OS effect has a dominant contribution to the energy shift. However, when $\delta\approx E_0$, the BS effect 
\cite{Bloch1940} causes similar energy shift as the OS effect. In the intermediate region of 
$0<\delta<E_0$, both effects are present and the ratio between the induced shifts is 
$\Delta E_{OS}/\Delta E_{BS}=(E_0+\hbar\omega)/(E_0-\hbar\omega)$ in the two-level approximation 
\cite{Sie2015,Sie2017}. Considering the selection rules in 2D TMDs for circularly-polarized resonant light, the two effects act separately in the two degenerate valleys in K$^+$ and K$^-$ points 
of the Brillouin zone generating an anisotropy of the exciton shifts in these two valleys. This effect can be applied, e.g., in ultrafast optical switches \cite{Gansen2002} or modulators \cite{Jin1990}.

Prior to our experimental observation of the valley-dependent OS and BS shifts only a few papers were devoted to these effects. These pioneering works, Refs.~[\onlinecite{Kim2014},\onlinecite{Sie2015},\onlinecite{Sie2017},\onlinecite{Cunningham2019},\onlinecite{LaMountain2018}], give basically a qualitative explanation of the observed Optical Stark and Bloch-Siegert shifts employing a phenomenological so-called two-level model.

In the present paper, the theoretical work is based on a model Hamiltonian comprising three salient components, the electron band structure, the Coulomb interaction and the coupling to external electromagnetic fields. The observed phenomena are described by the Semiconductor Bloch Equations (SBE), i. e., optical quantum transport equations. Even in the minimalistic version employed,is as most of the desired results are fully quantitative and can be obtained in a transparent analytical  form which can be back compared with the results of the two-level model.  

In our investigation we have focused on the improvements of the following limitations of the previous studies.   
First, none of the previous studies proposed an analytical expression for the transition dipole moment matrix elements. Only in one paper [\onlinecite{LaMountain2018}] these matrix elements were restored, as fitting parameters from the experiment and in two others papers [\onlinecite{Kim2014},\onlinecite{Cunningham2019}] the values, proportional to square of matrix elements, were derived from a similar fitting procedure. However, a full theoretical description should provide this number independently from the experiment.  
Second, the previous studies use a simplified two-level model for the explanation of the shifts. In particular, this phenomenological model doesn’t take into account the Coulomb many-body effects and effects of screening of the Coulomb interaction in TMD monolayers, which, as it turns out, are not negligible. 
Third, the previous studies didn’t provide the limits of applicability of their phenomenological description, e.g., at which intensity of the pump pulse the non-linear effects become comparable with the leading linear ones. 
Finally, the previous studies are focused predominantly on the A-exciton transitions, but the B-exciton transitions have not been discussed significantly except a brief study of the B-exciton OS shifts in 
Ref.~[\onlinecite{LaMountain2018}]. 

These four weak points of the previous studies motivated us to make a full theoretical investigation of the OS and BS effects and also perform the experiments which provide a) the data for B-excitons b) as high intensity of the pump pulse as possible to observe experimentally the limits of the proposed theory, and c) perform the experiment with two different TMD monolayers (WSe$_2$, MoS$_2$) in order to avoid accidental coincidence of experimental and theoretical results, i.e., to verify the correctness of our theoretical description for all TMD monolayers.

In this paper we study both theoretically and experimentally valley-selective blue shifts of 1sA and 1sB exciton resonances induced by off-resonant circularly-polarized optical fields in TMD crystals. We focus on WSe$_2$ and 
MoS$_2$ monolayers, which represent so-called darkish (with positive spin-splitting of the conduction band $\Delta_\text{c}>0$ in the K$^+$ point) and bright (with negative spin-splitting of the conduction band $\Delta_\text{c}<0$ 
in the K$^+$ point) materials, respectively \cite{Koperski2017}.
The theoretical predictions are compared with experimental results.
\begin{figure}[t]
	\centering
	\includegraphics[width=\linewidth]{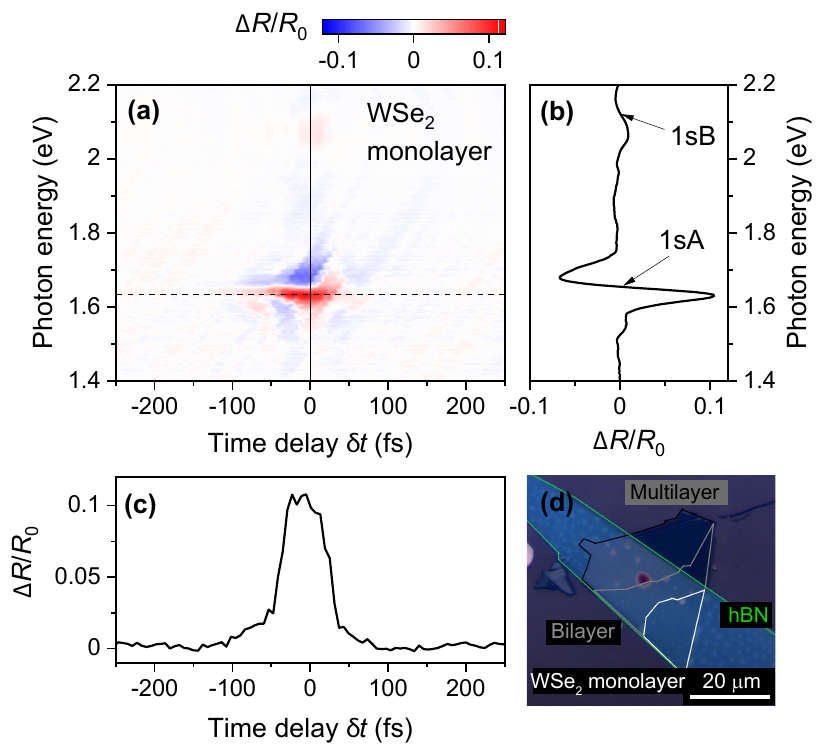}
	\caption{(a) Spectra of the transient reflectivity of the WSe$_2$ monolayer as a function of the time delay 
	between pump (photon energy of 0.62~eV, peak intensity of 17~GW/cm$^2$) and broadband probe pulses, both of the 
	same circular polarizations. (b) Transient reflectivity spectrum at zero time delay (solid line in (a)). (c) Time
	profile of the transient reflectivity signal at photon energy 1.63~eV (dashed line in (a)). (d) Microscope image of
	the investigated sample.}
	\label{fig:fig_1}
\end{figure}
\begin{figure}[t]
	\centering
	\includegraphics[width=\linewidth]{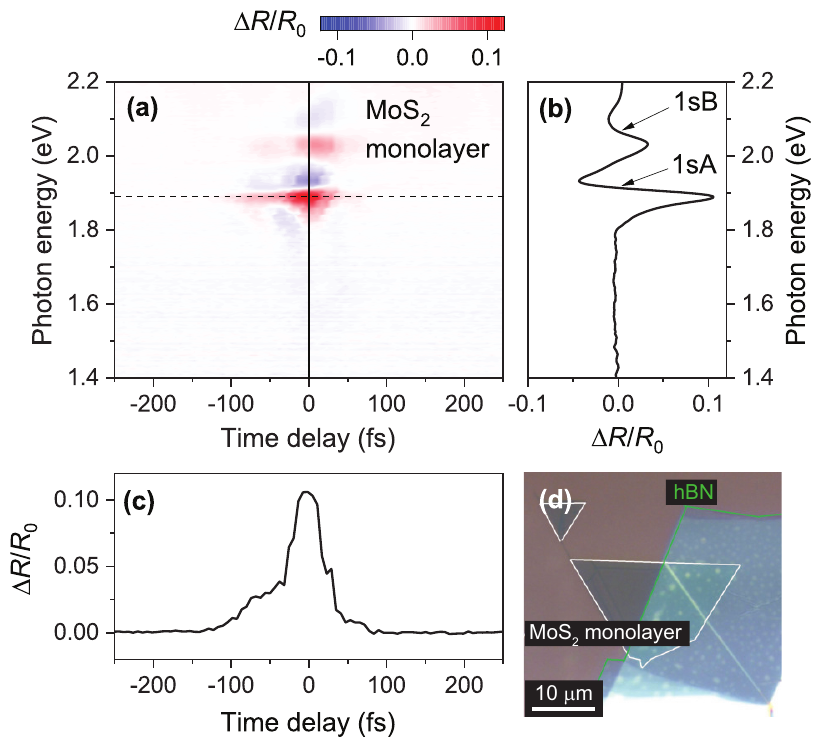}
	\caption{(a) Spectra of the transient reflectivity of the MoS$_2$ monolayer as a function of the time delay between pump (photon energy of 0.62~eV, peak intensity of 22~GW/cm$^2$) and broadband probe pulses, both of the same circular polarizations. (b) Transient reflectivity spectrum at zero time delay (solid line in (a)). (c) Time profile of the transient reflectivity signal at photon energy 1.89~eV (dashed line in (a)). (d) Microscope image of the investigated sample.}
	\label{fig:fig_2}
\end{figure}
\begin{figure*}[t]
	\centering
	\includegraphics[width=0.8\linewidth]{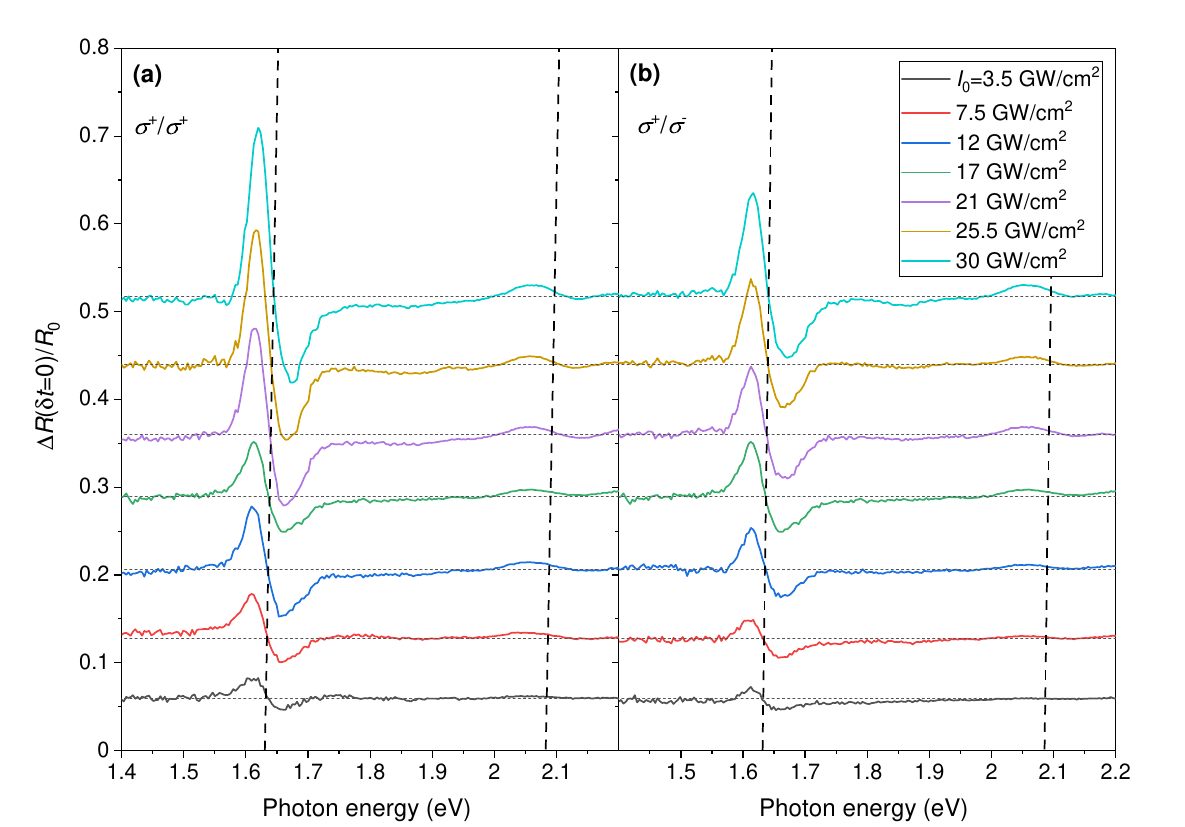}
	\caption{Spectra of the transient reflectivity change in WSe$_2$ monolayer at time delay $\delta t=0$ fs for (a) the same (optical Stark effect), $\tau=1$, and (b) opposite (Bloch-Siegert shift), $\tau=-1$, handednesses of circular polarizations of the pump and probe beams. Curves are vertically shifted for clarity by the distance proportional to the pump pulse intensity. The dashed lines show that the spectral shift increases linearly with the pump intensity (see the crossing points of the measured curves with the horizontal base lines indicating zero reflectivity change, which correspond to $E_0+\Delta E/2$). }
	\label{fig:fig_3}
\end{figure*}
\begin{figure*}[t]
	\centering
	\includegraphics[width=0.8\linewidth]{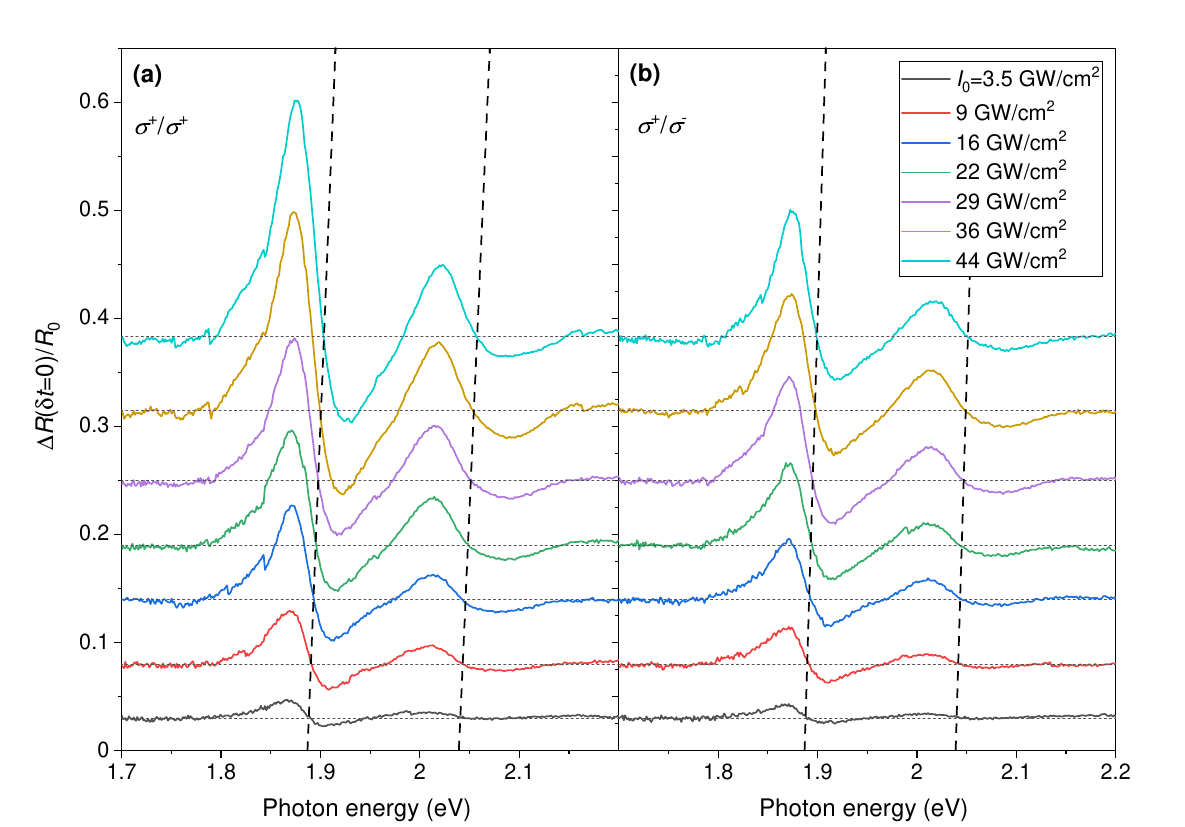}
	\caption{Spectra of the transient reflectivity change in MoS$_2$ monolayer at time delay $\delta t=0$ fs for (a) the same (OS effect), $\tau=1$, and (b) opposite (BS shift), $\tau=-1$, handednesses of circular polarizations of the pump and probe beams. Curves are vertically shifted for clarity by the distance proportional to the pump pulse intensity. The dashed lines show that the spectral shift increases linearly with the pump intensity (see the crossing points of the measured curves with the horizontal base lines indicating zero reflectivity change, which correspond to 
$E_0+\Delta E/2$).}
	\label{fig:fig_4}
\end{figure*}

The paper is organized as follows. In Sec.~\ref{sec:Experimental_results}, we provide the results of experimental measurements of the OS and BS shifts in WSe$_2$ and MoS$_2$ monolayers. 
In Sec.~\ref{sec:Semiconductor_Bloch_equations}, we present the theoretical model based on the semiconductor 
Bloch equations to explain the experimental results. In Sec.~\ref{sec:Polarization_dependent_optical_response},
we derive the analytical expressions for the OS and BS shifts in the studied monolayers. 
The numerical values of these shifts are calculated in Secs.~\ref{sec:Estimate_of_the_energy_shifts_for_WSe2} 
and \ref{sec:Estimate_of_the_energy_shifts_for_MoS2} for WSe$_2$ and MoS$_2$ monolayers, respectively and compared with the previously obtained results in Sec.~\ref{sec:other_results}.     
In Sec.~\ref{sec:Conclusions}, we summarize our results, discuss the advantages and limits of the proposed 
description of the excitonic shifts. Technical details are presented in Appendices 
\ref{app:pump_power}-\ref{app:estimation}.

\section{Experimental motivation}
\label{sec:Experimental_results}

To study the blue shift of the excitonic levels we perform transient reflectivity measurements of 2D TMD samples.
The samples are prepared by gel-film assisted mechanical exfoliation from bulk crystals. The monolayers are then transferred to a Si substrate with 90~nm thick layer of SiO$_2$ at the surface. Samples are covered by multilayer of hBN to protect them from degradation at ambient atmosphere (see Figs.~\ref{fig:fig_1}(d) and \ref{fig:fig_2}(d)).  

In the transient reflectivity experiments we measure the spectrum of relative reflectivity change  of a supercontinuum probe beam (photon energy 1.30-2.25~eV) as a function of the time delay with respect to the infrared pump pulse 
(central photon energy 0.62~eV, FWHM of the pulse duration of 38~fs). The arrival time of each spectral component of the broadband probe pulse with respect to the compressed pump pulse is measured using nonresonant nonlinearity in a thin glass. Time delays of spectral components are then shifted accordingly in the presented transient reflectivity data.

Both pulses are characterized by fixed circular polarizations and are focused using an off-axis parabolic mirror. An optical microscope setup is used to ensure the optimal focusing of the probe beam at the monolayers and to align the spatial overlap of the pump and probe beams. The circular polarizations of both pump and probe pulses are generated using broadband quarter-wave plates. 

All the measurements are carried out at room temperature with the laser repetition rate of 25 kHz. Due to the interference on the thin layer of SiO$_2$, the pump intensity in the monolayers is reduced to 0.35I$_0$, where I$_0$ is the peak intensity of the pump pulse in vacuum (see details in Appendix~\ref{app:pump_power}).

The results for the  transient reflectivity as a function of the photon energy of the probe pulse and the time delay for WSe$_2$ and MoS$_2$ monolayers are shown in Figs. \ref{fig:fig_1}(a) and \ref{fig:fig_2}(a).
The observed blue spectral shift of the exciton absorption peak manifests itself 
in the spectrum of $\Delta R(\hbar\omega,\delta t)/R_0(\hbar\omega)$. 
Here $\Delta R(\hbar\omega,\delta t)=R(\hbar\omega,\delta t)-R_0(\hbar\omega)$ is the 
difference between the transient reflectivities of the monolayer in the presence of the pump pulse 
$R(\hbar\omega,\delta t)$ and without it
$R_0(\hbar\omega)$. The parameter $\delta t$ defines the time delay between the pump and probe pulses. 
The procedure of evaluation of the reflectivity $R_0(\hbar\omega)$ of the monolayer on SiO$_2$/Si substrate is provided in Appendix~\ref{app:definition} (the same procedure is applied for determination of 
$R(\hbar\omega,\delta t)$).  

The position and width of the exciton resonances for each time delay $\delta t$ 
can be obtained from the analysis of the $\Delta R(\hbar\omega,\delta t)/R_0(\hbar\omega)$ in a frequency domain. 
Thus we clearly resolve features corresponding to the shift of 1sA and 1sB exciton resonances, see the spectra at zero time delay shown in Figs. \ref{fig:fig_1}(b) and \ref{fig:fig_2}(b).

When the excitonic shift $\Delta E$, induced by the pump pulse, is smaller than the width of the exciton peak, the reflectivity change in the frequency 
domain for a fixed delay time $\delta t$ can be approximated by the derivative of the spectral shape of the peak as  
\begin{align}
\frac{\Delta R(\hbar\omega,\delta t)}{R_0(\hbar\omega)}
\approx -\frac{\Delta E}{R_0(\hbar\omega)} \frac{dR_0(E)}{dE}\Big|_{E=\hbar\omega-\Delta E/2},
\end{align}
where we used the approximation $R(\hbar\omega,\delta t)\approx R_0(\hbar\omega-\Delta E)$ (see details in 
Supplemental Material of Ref.~[\onlinecite{PRL2022}]).
The amplitude of the reflectivity change thus scales linearly with the shift $\Delta E$. 
Such a behavior can be understood theoretically supposing that the influence of the pump pulse is parametrically small. However, at higher intensities of the pump pulse, the optical response of the sample can demonstrate features beyond the perturbation analysis. Below we demonstrate experimentally that the spectral changes become more complex at high pump intensities. 

In Figs. \ref{fig:fig_3} and \ref{fig:fig_4} we show the comparison of the transient reflectivity spectra in both samples measured at the time delay $\delta t=0$~fs (pump and probe pulses are overlapped) in the linear regime, in which the shift increases linearly with the peak intensity of the pump pulse. These data show a dependence of the amplitude of the signal and thus the amplitude of the shift on the combination of circular polarizations of the pump and probe beams. For the co-rotating polarizations corresponding to the OS effect we observe larger shifts while for the counter-rotating polarizations we observe smaller shifts caused by the BS effect. This is qualitatively in agreement with the two-level approximation which predicts the dependence 
$\Delta E_\mathrm{OS}=2|d_\mathrm{cv}|^2\mathcal{E}_0^2/(E_0-\hbar\omega)$ for the OS and 
$\Delta E_\mathrm{BS}=2|d_\mathrm{cv}|^2\mathcal{E}_0^2/(E_0+\hbar\omega)$ for the BS shifts 
(see details in \cite{Autler1955,Bloch1940,Sie2017} and Appendix~\ref{app:estimation_two_band}).
Here $d_\mathrm{cv}$  and $E_0$ are the transition dipole matrix element and the energy distance between the considered levels. $\mathcal{E}_0$ and $\omega $ are the amplitude of the electric field and the frequency of the pump pulse. However, it turns out that the coefficients of proportionality between $\Delta E_\mathrm{OS}$ and 
$|d_\mathrm{cv}|^2\mathcal{E}_0^2/(E_0-\hbar\omega)$ as well as between $\Delta E_\mathrm{BS}$ and 
$|d_\mathrm{cv}|^2\mathcal{E}_0^2/(E_0+\hbar\omega)$ are much larger than ``2'' due to the effects of the Coulomb interaction \cite{PRL2022}. This observation implies the limitations of the application of the two-level model for the quantitative estimation of the corresponding shifts in real materials and requires a more sophisticated analysis, e.g., semiconductor Bloch equations.        

\begin{figure*}[t]
	\centering
	\includegraphics[width=0.8\linewidth]{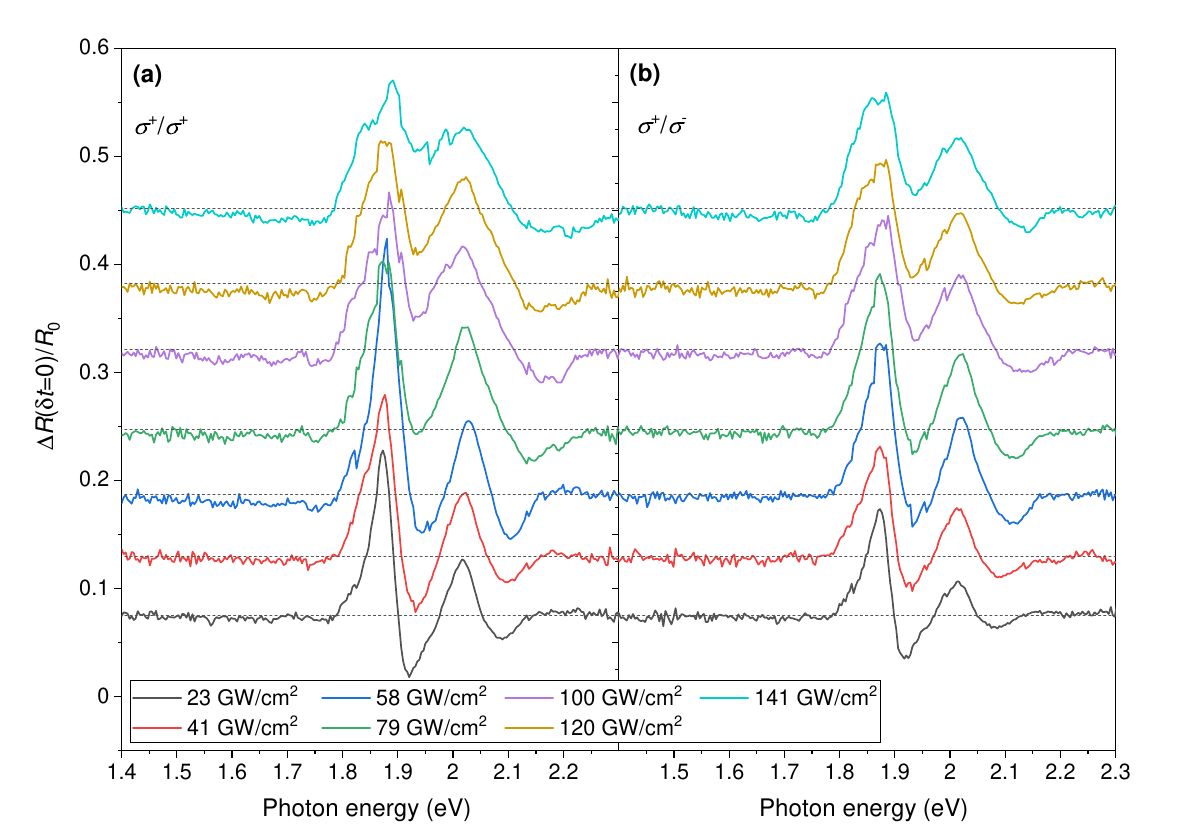}
	\caption{Spectra of the transient reflectivity change in MoS$_2$ monolayer at time delay $\delta t=0$ fs for (a) the same (optical Stark effect), $\tau=1$, and (b) opposite (Bloch-Siegert shift), $\tau=-1$, handednesses of circular polarizations of the pump and probe beams in the nonlinear strong-field regime. Curves are vertically shifted for clarity by the distance proportional to the pump pulse intensity.}
	\label{fig:fig_5}
\end{figure*}

\begin{figure}[t]
	\centering
	\includegraphics[width=\linewidth]{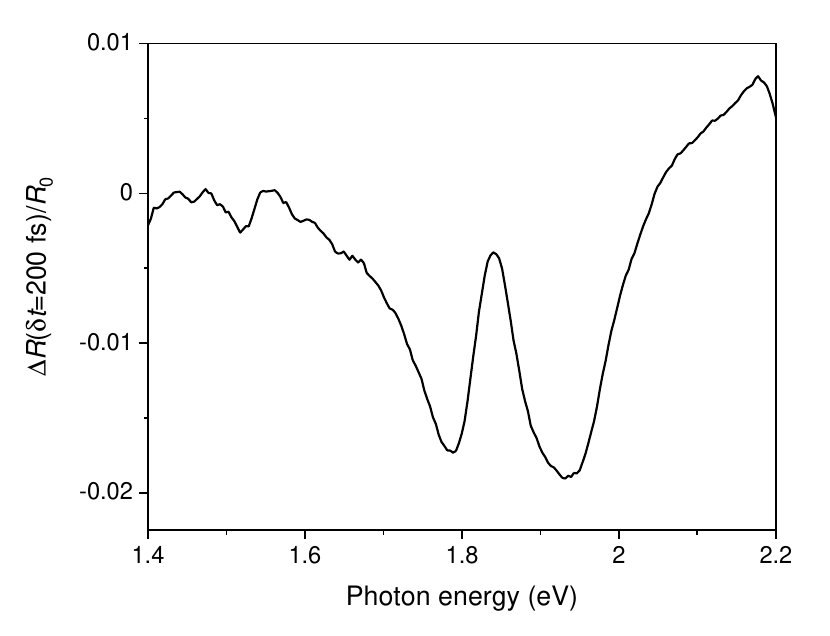}
	\caption{The transient reflectivity spectra in MoS$_2$ monolayer at the time delay 200~fs after the excitation.}
	\label{fig:fig_6}
\end{figure}

When the peak intensity of the pump pulse overcomes certain threshold, the transient reflectivity spectrum changes its shape. We show the transition from the linear to the nonlinear regime in Fig. \ref{fig:fig_5}, where the transient reflectivity spectra measured in the MoS$_2$ monolayer at the time delay $\delta t=0$~fs are compared for both combinations of circular polarizations of the pump and probe pulses. At intensities above 50 GW/cm$^2$, the pump excites real population of carriers via multiphoton absorption. The exciton transitions clearly broaden due to the exciton-exciton interaction. The population of real excitons is also visible at longer time delays via the bleaching of absorption of the probe pulse at the resonant frequency. This can be seen in Fig. \ref{fig:fig_6}, where we show the transient reflectivity spectra at the time delay 200~fs after the excitation.
Another feature, which appears in the transient spectra at high intensities of the pump, is a decrease of reflectivity (increase of absorption) at photon energies below the band gap. The origin of this decrease may be related to entering the strong-field regime of the interaction, which is related to the onset of the dynamical Franz-Keldysh effect (DFKE) induced by the pump pulse. This effect causes a blue shift of the band gap and an increase of absorption at photon energies below the band gap. It was predicted and observed in several semiconductors \cite{Jauho1996,Srivastava2004} and in the excitonic system in quantum wells \cite{Nordstrom1998}, which were illuminated by strong THz electromagnetic fields. DFKE is observable when the ponderomotive energy of the electron-hole pairs $U_p$ becomes comparable to the photon energy of the driving wave 
$\hbar\omega$ \cite{Nordstrom1998}. The ponderomotive energy is defined as the time-averaged value of the kinetic energy of the particle in the oscillating electric field of the pump, which for circularly polarized light reads 
$U_p=e^2\mathcal{E}_0^2/(2m\omega^2)$, where $m$ is the reduced mass of the exciton. The maximum pump intensity applied to the monolayer in our experiments corresponds to the ratio $\gamma=U_p/(\hbar\omega)\approx 0.25$ which is close to the value of $\gamma=1$, which is characteristic for the DFKE.

\section{Semiconductor Bloch equations}
\label{sec:Semiconductor_Bloch_equations}

The goal of this section is to propose a theoretical model which explains the experimental observations:
{\it i)} the shift of the exciton energy in the presence of the strong off-resonant pump pulse; 
{\it ii)} the linear scaling of the shift with the intensity of the pump pulse; 
{\it iii)} the dependence of the shift on the handedness of the circular polarization of the pump pulse. 
According to the first statement, the optical response of the monolayer in the presence of the strong pump pulse
is determined by the energies of the excitons and not by the energies of interband transitions.
Therefore the Coulomb interaction, which is responsible for formation of the excitons, must be included in the model. 
The second statement claims that the system remains in a linear response regime, 
i.e., the polarization $\mathbf{P}$ induced in monolayer by the electric field $\mathbf{E}$ of pump pulse
scales linearly with the electric field. It gives the following qualitative estimate of the energy shift 
in the system $\Delta E\propto \mathbf{P\cdot E}\propto\chi_{ij}E_iE_j\propto |\mathbf{E}|^2$, where $\chi_{ij}$ 
represents symbolically the susceptibility matrix. It allows to consider the influence 
of the pump pulse perturbatively in the $|\mathbf{E}|^2$ parameter.  
Finally, the third statement implies an important role of the optical selection rules to explain the observed 
OS and BS effects in TMD monolayer.    
 
\begin{figure}[t]
	\centering
	\includegraphics[width=0.8\linewidth]{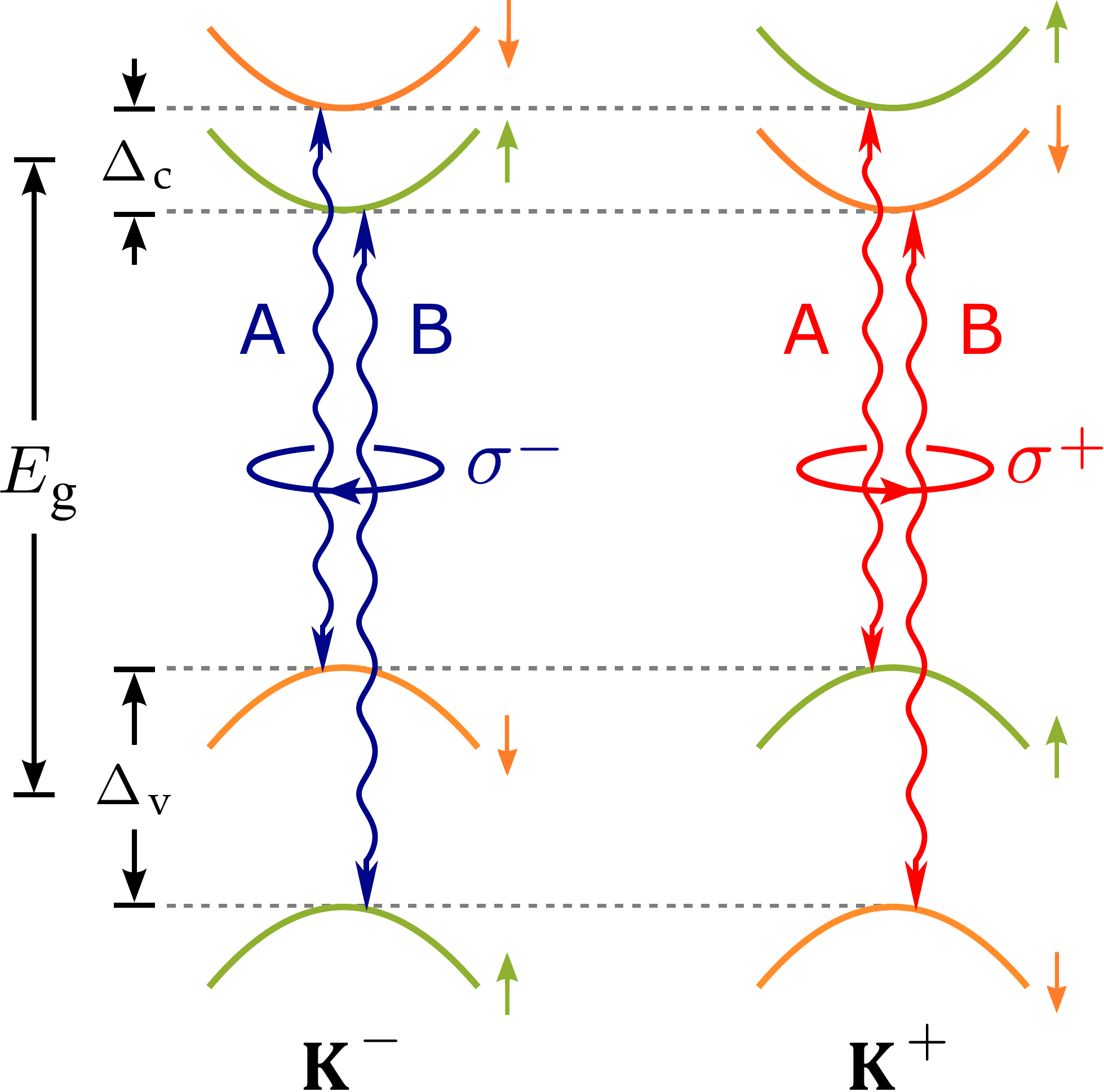}
	\caption{Spin and valley structure of WSe$_2$ monolayer. Green and orange solid curves show 
	spin-up and spin-down bands in the K$^\pm$ valleys, respectively. $\Delta_\text{c}$ and $\Delta_\text{v}$ are the 
	splittings of the conduction and valence bands, and $E_\text{g}$ is the single particle band-gap in the system. 
	Double-headed wavy arrows represent spin-allowed excitonic transitions in each valley. The blue/red color of the
	corresponding arrows indicate the left/right circular polarization of light which couples electromagnetically the
	corresponding bands. The capital letters A/B label the low-energy/high-energy excitonic transitions in each valley. }
	\label{fig:fig_0}
\end{figure}

Note that optical transitions between spin-up (spin-down) valence and conduction bands in K$^+$ (K$^-$) valley
lead to the formation of intravalley A excitons (see Fig.~\ref{fig:fig_0}). 
Optical transitions between spin-down (spin-up) valence and conduction bands in K$^+$ (K$^-)$ valley 
lead to the formation of intravalley B excitons. 
The A and B exciton transitions of the same (opposite) valley do not affect each other due to the spin (momentum) conservation selection rule. It allows to consider each type of excitons (A exciton in $\tau=1$ valley, 
A exciton in $\tau=-1$ valley, B exciton in $\tau=1$ valley, 
B exciton in $\tau=-1$ valley) separately (see Fig.~\ref{fig:fig_0}).
Therefore, one needs to consider 4 cases, for each pair of valence and conduction bands where optical 
transitions happen (with the same spin and in the same valley) separately. 
It turns out that the quasiparticle Hamiltonians for each aforementioned case have the same structure, 
therefore we consider them uniformly for brevity.    
 
The two-band second quantized quasiparticle Hamiltonian in $\tau$ valley reads 
\begin{equation}
H_\text{b}^\tau=\sum_\mathbf{k} E_{e,k}\alpha^{\tau\dag}_\mathbf{k}\alpha^\tau_\mathbf{k}+
E_{h,k}\beta^{\tau\dag}_{-\mathbf{k}}\beta^\tau_{-\mathbf{k}}.
\end{equation} 
Here $E_{e,k}=\hbar^2k^2/2m_e+\widetilde{E}_\mathrm{g}$ and $E_{h,k}=\hbar^2k^2/2m_h$ are 
the dispersion of electrons and holes for the chosen pair of the conduction and valence bands,
respectively, and $k=|\mathbf{k}|$. 
$m_e,m_h>0$ are the electron and hole effective masses of the considering bands, 
$\widetilde{E}_\mathrm{g}$ is the bandgap in the system, $\alpha^\tau_\mathbf{k}$ and 
$\beta^\tau_{-\mathbf{k}}$ are the annihilation 
operators for electrons and holes, with momentum $\mathbf{k}$ in $\tau$ valley of the 
considering bands.
The parameters $m_e,m_h, \widetilde{E}_\text{g}$ are different for the case of A and B exciton transitions. 
However these parameters are the same for the A(B) exciton transitions in different valleys. 
The latter is the consequence of the time-reversal symmetry of TMD crystals.
     
The Coulomb interaction in the system is given by the Hamiltonian
\begin{align}
H_\text{C}^\tau&=
\sum_{\mathbf{k},\mathbf{k}',\mathbf{q}\neq 0}\!\!\!\frac{V_\mathbf{q}}{2} (\alpha^{\tau\dag}_{\mathbf{k}+\mathbf{q}}
\alpha^{\tau\dag}_{\mathbf{k}'-\mathbf{q}}\alpha^\tau_{\mathbf{k}'}\alpha^\tau_\mathbf{k}+
\beta^{\tau\dag}_{\mathbf{k}+\mathbf{q}}\beta^{\tau\dag}_{\mathbf{k}'-\mathbf{q}}
\beta^\tau_{\mathbf{k}'}\beta^\tau_\mathbf{k})\nonumber \\
&-\sum_{\mathbf{k},\mathbf{k}',\mathbf{q}\neq 0}V_\mathbf{q} 
\alpha^{\tau\dag}_{\mathbf{k}+\mathbf{q}}
\beta^{\tau\dag}_{\mathbf{k}'-\mathbf{q}}\beta^\tau_{\mathbf{k}'}\alpha^\tau_\mathbf{k},
\end{align}
where $V_\mathbf{q}$ is the Fourier transform of the Rytova-Keldysh potential 
\cite{Rytova1967,Keldysh1979,Cudazzo2011}. The first line in the Hamiltonian 
describes the electron-electron and hole-hole Coulomb repulsion, while the second line
describes the electrons-hole Coulomb interaction and leads to formation 
of bright A (B) excitons in the system. Note that we don't include the Coulomb interaction 
between the quasiparticles from different valleys in our consideration. 
Such terms are responsible for the exchange interaction between bright excitons of opposite 
valleys \cite{Glazov2014,Gartstein2015,Slobodeniuk2016}. The effects of exchange interaction 
are much smaller than the effects considered in the current study and hence can be neglected. 
Therefore, the Coulomb interaction in the system splits into a sum of independent terms 
by valley and by bands spin indices.      
Then, the full Hamiltonian for $\tau$ valley in the absence of the external fields is the sum 
of the band and Coulomb Hamiltonians $H^\tau=H^\tau_\text{b}+H^\tau_\text{C}$.  

The interaction of the monolayer with $\sigma^\pm$ polarized light is defined as 
$H^\tau_\text{int}=-\mathbf{P}^\tau\cdot\mathbf{E}$. Here $\mathbf{P}^\tau$ is the polarization 
operator of the system in $\tau$ valley, 
and $\mathbf{E}=\mathcal{E}(t)[\mathbf{e}_x\cos(\omega t)\pm \mathbf{e}_y\sin(\omega t)]$ 
is the electric field of the $\sigma^\pm$ polarized light with normal incidence. Note that $\mathcal{E}(t)$ is the amplitude of the electric field of light. The time-independent amplitude $\mathcal{E}(t)=\mathcal{E}$ corresponds to the case of the monochromatic plane wave. The light-matter interaction Hamiltonian in the second quantized form reads 
\begin{equation}
H^\tau_\text{int}=-\sum_\mathbf{k} 
d^\tau_\mathrm{cv}\mathcal{E}^\tau_\pm(t)\alpha^{\tau\dag}_\mathbf{k}\beta^{\tau\dag}_{-\mathbf{k}} + \text{h.c.}.
\end{equation}			
Here $d_\mathrm{cv}^\tau=\tau d_\mathrm{cv}$ is the transition dipole moment between the valence and conduction bands
and $\mathcal{E}^\tau_\pm(t)=\mathcal{E}(t)\exp(\mp i\tau\omega t)$. 
Note that $H^\tau_\text{int}$ has a similar form as at for the two-level system in 
the rotating-wave approximation. However, such a form of $H^\tau_\text{int}$ originates from the specific 
structure of the interband transition dipole moments in $\mathrm{K}^\pm$ points of TMDs, and no additional restrictions on the frequencies of the pulse are implied (see details in Appendices~\ref{app:two_band_model} and 
\ref{app:quantum_interaction}).  
 
We use the Semiconductor Bloch equations (SBE) approach to evaluate the energy shift of the exciton transitions 
in the monolayer. To do so we consider the quantum average of the polarization 
$P^\tau_\mathbf{k}(t)=\langle\beta^\tau_{-\mathbf{k}}\alpha^\tau_\mathbf{k}\rangle$, 
and the electrons $n^\tau_{\mathbf{k},e}(t)=\langle\alpha_\mathbf{k}^{\tau\dag}\alpha^\tau_\mathbf{k}\rangle$ and holes $n^\tau_{\mathbf{k},h}(t)=\langle\beta^{\tau\dag}_{-\mathbf{k}}\beta^\tau_{-\mathbf{k}}\rangle$
populations. The SBE read \cite{Haug2009}
\begin{align}
\label{eq:bloch_polarization}
\frac{\partial P^\tau_\mathbf{k}}{\partial t}&=-ie^\tau_kP^\tau_\mathbf{k}-
i(n^\tau_{\mathbf{k},e}+n^\tau_{\mathbf{k},h}-1)\omega^\tau_{R,\mathbf{k}}+ 
\frac{\partial P^\tau_\mathbf{k}}{\partial t}\Big|_{scatt},\\
\label{eq:bloch_electrons}
\frac{\partial n^\tau_{\mathbf{k},e}}{\partial t}&=
i(\omega^\tau_{R,\mathbf{k}}P^{\tau*}_\mathbf{k}-\omega^{\tau*}_{R,\mathbf{k}}P^\tau_\mathbf{k})+
\frac{\partial n^\tau_{\mathbf{k},e}}{\partial t}\Big|_{scatt},\\
\label{eq:bloch_holes}
\frac{\partial n^\tau_{\mathbf{k},h}}{\partial t}&=
i(\omega^\tau_{R,\mathbf{k}}P^{\tau*}_\mathbf{k}-\omega^{\tau*}_{R,\mathbf{k}}P^\tau_\mathbf{k})+
\frac{\partial n^\tau_{\mathbf{k},h}}{\partial t}\Big|_{scatt}
\end{align} 
where we introduced the effective energy parameter 
\begin{equation}
\label{eq:energy_parameter}
\hbar e^\tau_k=\widetilde{E}_\mathrm{g}+\hbar^2k^2/2m-
2\sum_\mathbf{q} V_{\mathbf{k}-\mathbf{q}}n^\tau_\mathbf{q},
\end{equation} 
with the exciton reduced mass $m=m_em_h/(m_e+m_h)$,
and the Rabi frequency $\omega^\tau_{R,\mathbf{k}}$, with  
\begin{equation}
\label{eq:Rabi_frequency}
\hbar\omega^\tau_{R,\mathbf{k}}=d^\tau_\mathrm{cv}\mathcal{E}^\tau_\pm(t)+
\sum_{\mathbf{q}\neq \mathbf{k}}V_{\mathbf{k}-\mathbf{q}}P^\tau_\mathbf{q}.
\end{equation}
The last terms in Eqs.~(\ref{eq:bloch_polarization})-(\ref{eq:bloch_holes}) represent the dissipative
processes in the form of dephasing rates for the interband polarization $P^\tau_\mathbf{k}$ and collision rates 
for the electron $n^\tau_{\mathbf{k},e}$ and hole $n^\tau_{\mathbf{k},h}$ populations. 
The dissipative terms include effects of carrier-phonon interaction, carrier-carrier and 
carrier-impurities scattering, and the correlations effects beyond the Hartree-Fock approximation.
Moreover, these terms also can contain the processes which effectively couple the electronic states of 
opposite valleys, e.g., X- and Y-processes \cite{Schmidt2016} or higher-order Coulomb correlation effects like biexcitons \cite{Katsch2020,Selig2020}. 

Unfortunately, scattering terms make the analytical investigation of the SBE in a general form practically impossible. In order to simplify this problem we make the following assumptions. First, we limit our study to pure TMD crystals to exclude the impurity-induced intra- and intervalley 
scattering effects. Second, we take into account the peculiarities of our experiment, where strong non-resonant pump and weak resonant probe pulses are applied to the monolayer. The role of each pulse is different. The former pulse provides a renormalized ground state of the S-TMD monolayer, while the linear response to the latter (on the background of the modified ground state) yields the corresponding renormalized spectrum of excitons, which exhibits the OS and BS effects \cite{Chemla1989,SchmittRink1988_1,SchmittRink1988}.  
  
The non-resonant pump pulse induces only ``virtual electron-hole pairs'' which are characterized by polarization 
$P_\mathbf{k}^\tau$ and occupations $n^\tau_{\mathbf{k},e}$, $n^\tau_{\mathbf{k},h}$. 
They are responsible for the effective increase of the energy distance between valence and conduction band in each
$\tau=\pm 1$ valley, which leads to OS and BS shifts. This process is coherent and isn't characterized by noticable dissipation (see discussion in Refs.~[\onlinecite{Chemla1989},\onlinecite{SchmittRink1988},\onlinecite{Hemla1986},\onlinecite{Zimmermann1990}]). Our experimetal results presented in 
Figs.~\ref{fig:fig_1}(c) and \ref{fig:fig_2}(c) confirm this observation -- the effect of the applied non-resonant pump pulse, manifested in the OS and BS shifts, persists only when the pump pulse is present. This indicates that the shifts arise from coherent light-matter interaction between the pump pulse and the monolayer, rather than incoherent processes related to photoexcited excitons and/or charge carriers. 
Therefore, we exclude the scattering terms associated with the pump-field-induced quantities 
$P_\mathbf{k}^\tau$, $n^\tau_{\mathbf{k},e}$, $n^\tau_{\mathbf{k},h}$. Note that the statement about the coherent nature of the light-matter interaction has been exploited in the previous works, however the duration of their pump pulse was an order of magnitude larger (250 fs \cite{Kim2014}, 160 fs \cite{Sie2015,Sie2017}, 150 fs \cite{Cunningham2019}, 375 fs \cite{LaMountain2018}) than in our experiment. 

The resonant probe pulse generates the real electron-hole pairs, which correspond to the polarizations 
$\delta P_\mathbf{k}^\tau$ and electron/hole $\delta n^\tau_{\mathbf{k},e/h}$ occupation numbers. These excitons can decay via various processes, which together define the width of the exciton line of tens of meV (or equivalently hundred of fs) at room temperature. The simplest phenomenological way to include this relaxation process into account 
is to introduce the dephasing $\gamma$ and decoherence $\Gamma$ rates into 
Eqs.~(\ref{eq:bloch_polarization})-(\ref{eq:bloch_holes}) with the following substitution 
$(\partial P^\tau_\mathbf{k}/\partial t)|_{scatt}\rightarrow -\gamma \delta p_\mathbf{k}^\tau$, 
$(\partial n^\tau_{\mathbf{k},e}/\partial t)|_{scatt}\rightarrow -\Gamma\delta n^\tau_{\mathbf{k},e}$, 
$(\partial n^\tau_{\mathbf{k},h}/\partial t)|_{scatt}\rightarrow -\Gamma \delta n^\tau_{\mathbf{k},h}$. 
In the latter we suppose that the effective relaxation rates for electron and holes have the same values. 

Then, introducing the decaying terms in Eqs. ~(\ref{eq:bloch_polarization})-(\ref{eq:bloch_holes}), 
we write the SBE for full polarization 
$P_\mathbf{k}^\tau+\delta p_\mathbf{k}^\tau$ and occupation numbers 
$n^\tau_{\mathbf{k},e/h}+\delta n_{\mathbf{k},e/h}^\tau$. Taking into account that 
$|P_\mathbf{k}^\tau|,|n^\tau_{\mathbf{k},e/h}|\gg |\delta p_\mathbf{k}^\tau|,|\delta n^\tau_{\mathbf{k},e/h}|$
we derive the hierarchy of the equations considering the values 
$\delta p_\mathbf{k}^\tau,\delta n^\tau_{\mathbf{k},e/h}$ as small parameters.   
In the leading order we obtain the equation for $P_\mathbf{k}^\tau$ and $n^\tau_{\mathbf{k},e/h}$, generated by the pump pulse only. The solutions of these equations (see Appendix~\ref{app:polarization}) we substitute to the next group of the equations from the hierarchy.  
The new equations will contain linear in $\delta p_\mathbf{k}^\tau$ terms (together with the relaxation one) and 
the pump pulse induced $P^\tau_\mathbf{k}$ and $n^\tau_{\mathbf{k},e/h}$ dependent part. 

The qualitative analysis of this equation claims that the relaxation terms are responsible only for the broadening of the excitonic lines and don't influence the excitonic shifts, if $\gamma,\Gamma$ are much smaller than the photon energies of the pump and probe pulses (see discussion in the Sec.~\ref{sec:Polarization_dependent_optical_response}). 
On the other hand, the $P^\tau_\mathbf{k}$ and $n^\tau_{\mathbf{k},e/h}$ dependent part provides 
the many-body Coulomb interaction correction to the excitonic shifts.  
Since in this study we are focused on the excitonic shifts and don't study the effects of broadening of the 
excitonic lines it will be enough to consider the case $\gamma=\Gamma=0$ as a starting point. 
This case corresponds to infinitesimally narrow excitonic lines. 

Summarising the results of our analysis we conclude that the excitonic shifts in TMD monolayers 
for our particular problem can be obtained from the SBE (\ref{eq:bloch_polarization})-(\ref{eq:bloch_holes}) 
without scattering terms 
(where we made a replacement $P_\mathbf{k}^\tau+\delta p_\mathbf{k}^\tau\rightarrow P_\mathbf{k}^\tau$ and  
$n^\tau_{\mathbf{k},e/h}+\delta n_{\mathbf{k},e/h}^\tau\rightarrow n^\tau_{\mathbf{k},e/h}$ in the SBE 
for brevity). Note that, the Eqs.~(\ref{eq:bloch_electrons}) and (\ref{eq:bloch_holes}) coincide  
in this case leading to the additional relation
\begin{equation}
\frac{\partial (n^\tau_{\mathbf{k},e}-n^\tau_{\mathbf{k},h})}{\partial t}=0
\end{equation}
Hence, the difference of electrons and holes populations doesn't depend on time. 
Taking into account the electroneutrality of the crystal at the initial moment of time $t_{in}$,
$n^\tau_{\mathbf{k},e}(t_{in})=n^\tau_{\mathbf{k},h}(t_{in})=0$, we conclude that 
$n^\tau_{\mathbf{k},e}=n^\tau_{\mathbf{k},h}\equiv n^\tau_\mathbf{k}$.

The system of SBE equations (\ref{eq:bloch_polarization})-(\ref{eq:bloch_holes}) then reads
\begin{align}
\label{eq:bloch_p}
\frac{\partial P^\tau_\mathbf{k}}{\partial t}=&-ie^\tau_kP^\tau_\mathbf{k}-
i(2n^\tau_\mathbf{k}-1)\omega^\tau_{R,\mathbf{k}},\\
\label{eq:bloch_n}
\frac{\partial n^\tau_\mathbf{k}}{\partial t}=&
i(\omega^\tau_{R,\mathbf{k}}P^{\tau*}_\mathbf{k}-\omega^{\tau*}_{R,\mathbf{k}}P^\tau_\mathbf{k}).
\end{align} 
It contains an integral of motion  
\begin{equation}
(1-2n_\mathbf{k}^\tau)^2+4|P_\mathbf{k}^\tau|^2=1,
\end{equation}
which can be verified by taking the time derivative of the expression on the l.h.s with further substituting  
the corresponding derivatives from the Eqs.~(\ref{eq:bloch_p}) and (\ref{eq:bloch_n}). 
It means that $n_\mathbf{k}^\tau$ and $P_\mathbf{k}^\tau$ variables are not independent, 
and one can be expressed as a function of the other one as
\begin{equation}
n_\mathbf{k}^\tau=\frac12(1-\sqrt{1-4|P_\mathbf{k}^\tau|^2}). 
\end{equation}
For the case of small $P_\mathbf{k}^\tau\ll1$ we have an approximate expression 
\begin{equation}
n_\mathbf{k}^\tau\approx |P_\mathbf{k}^\tau|^2, 
\end{equation}
which will be used further for the perturbative analysis and solution of the non-linear equations 
(\ref{eq:bloch_p}) and (\ref{eq:bloch_n}). 

\section{Polarization dependent optical response}
\label{sec:Polarization_dependent_optical_response}

We consider the two pulse experiment, where the pump (p) and probe/test (t) pulses are applied to the monolayer. There 
are 4 different possible combinations of their circular polarizations: $\sigma_\text{p}^+/\sigma_\text{t}^+$, 
$\sigma_\text{p}^+/\sigma_\text{t}^-$, $\sigma_\text{p}^-/\sigma_\text{t}^+$, 
$\sigma_\text{p}^-/\sigma_\text{t}^-$. 
Due to the time reversal symmetry the optical response of the monolayer in the $\tau$ valley 
for the case $\sigma_\text{p}^+/\sigma_\text{t}^\pm$ is equal to the 
the optical response of the monolayer in the $-\tau$ valley 
for the case $\sigma_\text{p}^-/\sigma_\text{t}^\mp$. 
This statement is also verified experimentally, see Appendix~\ref{app:verification}.    
Therefore it is enough to consider only the first pair of polarizations 
$\sigma_\text{p}^+/\sigma_\text{t}^+$, $\sigma_\text{p}^+/\sigma_\text{t}^-$ 
for each valley of the monolayer $\tau=\pm 1$.      

\subsection{$\sigma_\text{p}^+/\sigma_\text{t}^+$ case, $\tau=1$}
\label{subsec:tau_plus}
This case corresponds to the optical response of the crystal in the K$^+$ point. 
The dipole moment matrix element is $d^\tau_\mathrm{cv}=d_\mathrm{cv}$ and 
electric field of the pulses reads
\begin{equation}
\mathcal{E}=\mathcal{E}_\text{p}e^{-i\omega_\text{p}t}+\mathcal{E}_\text{t}e^{-i\omega_\text{t}t}, 
\end{equation}
where we have introduced the amplitudes $\mathcal{E}_\text{p},\mathcal{E}_\text{t}$ and frequencies 
$\omega_\text{p},\omega_\text{t}$ of the pump and test beams. 
Note that we approximate the time-dependent amplitude of the pump pulse by its average value
over the time of the pulse duration. It simplifies the Bloch equations without losing 
the effect of exciton energy shifts observed in the experiment. 
We take into account that 
$|\mathcal{E}_\text{p}|\gg|\mathcal{E}_\text{t}|$. 
Hence the pump pulse becomes the main source of the polarization $P_\mathbf{k}$ and concentration 
$n_\mathbf{k}$ in the system, while the test pulse generates only 
small perturbations $\delta P_\mathbf{k}$, $\delta n_\mathbf{k}$. 
 
Using the substitution $P_\mathbf{k}+\delta P_\mathbf{k}$ and $n_\mathbf{k}+\delta n_\mathbf{k}$ 
we linearize the equation of motion~(\ref{eq:bloch_p}) 
\begin{align}
i\frac{\partial \delta P_\mathbf{k}}{\partial t}=\delta e_kP_\mathbf{k}+e_k\delta P_\mathbf{k}+
(2n_\mathbf{k}-1)\delta\omega_{R,\mathbf{k}}+2\delta n_\mathbf{k}\omega_{R,\mathbf{k}}.
\end{align} 
Here
$\delta e_k=-(2/\hbar)\sum_{\mathbf{k}'}V_{\mathbf{k}-\mathbf{k}'}\delta n_{\mathbf{k}'}$
is derived from the definition of $e_k$,
$\delta n_\mathbf{k}=(P_\mathbf{k}\delta P_\mathbf{k}^*+\delta P_\mathbf{k} P_\mathbf{k}^*)/(1-2n_\mathbf{k})$
is a consequence of the integral of motion, and 
\begin{equation}
\hbar \delta \omega_{R,\mathbf{k}}=d_\mathrm{cv}\mathcal{E}_\text{t}e^{-i\omega_\text{t} t}+
\sum_{\mathbf{k}'\neq \mathbf{k}}V_{\mathbf{k}-\mathbf{k}'}\delta P_{\mathbf{k}'}.
\end{equation}
One can see that we can eliminate the time dependence of the pump field supposing that 
$P_\mathbf{k}=p_\mathbf{k}e^{-i\omega_\text{p}t}$, 
$\delta P_\mathbf{k}=\delta p_\mathbf{k}e^{-i\omega_\text{p}t}$. Then, taking into account that 
$\delta n_\mathbf{k} \sim 1$, $\omega_{R,\mathbf{k}}\sim e^{-i\omega_\text{p}t}$ and 
redefining the Rabi frequencies 
\begin{align}
\hbar \omega_{R,\mathbf{k}}\rightarrow& \hbar \omega_{R,\mathbf{k}}=
d_\mathrm{cv}\mathcal{E}_\text{p}+\sum_{\mathbf{k}'\neq \mathbf{k}}V_{\mathbf{k}-\mathbf{k}'}p_{\mathbf{k}'}, \\
\hbar \delta \omega_{R,\mathbf{k}}\rightarrow& \hbar \delta \omega_{R,\mathbf{k}}=
d_\mathrm{cv}\mathcal{E}_\text{t}e^{i\Delta t}+\sum_{\mathbf{k}'\neq \mathbf{k}}V_{\mathbf{k}-\mathbf{k}'}
\delta p_{\mathbf{k}'},
\end{align} 
where $\Delta=\omega_\text{p}-\omega_\text{t}$, we get the following equation 
\begin{align}
\label{eq:delta_p}
i\frac{\partial \delta p_\mathbf{k}}{\partial t}=&\delta e_kp_\mathbf{k}+(e_k-\omega_\text{p})\delta p_\mathbf{k}+ 
\nonumber \\ +&
(2n_\mathbf{k}-1)\delta\omega_{R,\mathbf{k}}+2\delta n_\mathbf{k}\omega_{R,\mathbf{k}}.
\end{align}
We are looking for the solutions of this equation in the form $\delta p_\mathbf{k}=a_\mathbf{k}e^{i\Delta t}+
b_\mathbf{k}e^{-i\Delta t}$. 
Substituting it into the equation and separating the positive $\sim e^{i\Delta t}$ and negative $\sim e^{-i\Delta t}$
frequency solutions we get the following set of equations 
\begin{align}
\sum_{\mathbf{k}'}\Big[&H^0_{\mathbf{k}\mathbf{k}'}+\delta H_{\mathbf{k}\mathbf{k}'}-
\hbar\omega_\text{t}\delta_{\mathbf{k}\mathbf{k}'}\Big]a_{\mathbf{k}'}=
(1-2|p_\mathbf{k}|^2)d_\mathrm{cv}\mathcal{E}_\text{t}+
\nonumber \\+& 
2\sum_{\mathbf{k}'}V_{\mathbf{k}-\mathbf{k}'}p_\mathbf{k}p_{\mathbf{k}'}(b_{\mathbf{k}'}^*-b_\mathbf{k}^*)-
2p_\mathbf{k}d_\mathrm{cv}\mathcal{E}_\text{p}b_\mathbf{k}^*,\\
\sum_{\mathbf{k}'}\Big[&H^0_{\mathbf{k}\mathbf{k}'}+\delta H_{\mathbf{k}\mathbf{k}'}+
\hbar(\omega_\text{t}-2\omega_\text{p})\delta_{\mathbf{k}\mathbf{k}'}\Big]b_{\mathbf{k}'}= \nonumber \\=&
2\sum_{\mathbf{k}'}V_{\mathbf{k}-\mathbf{k}'}p_\mathbf{k}p_{\mathbf{k}'}(a_{\mathbf{k}'}^*-a_\mathbf{k}^*)-
2p_\mathbf{k}d_\mathrm{cv}\mathcal{E}_\text{p}a_\mathbf{k}^*,
\end{align}
where we have introduced 
\begin{equation}
\label{eq:h_0}
H^0_{\mathbf{k}\mathbf{k}'}\equiv\Big(\widetilde{E}_\mathrm{g}+\frac{\hbar^2k^2}{2m}\Big)\delta_{\mathbf{k}\mathbf{k}'}-
V_{\mathbf{k}-\mathbf{k}'},
\end{equation}
\begin{align}
\label{eq:hkk_plus}
\delta H_{\mathbf{k}\mathbf{k}'}\equiv&
2\delta_{\mathbf{k}\mathbf{k}'}\Big[\sum_{\mathbf{k}''}V_{\mathbf{k}-\mathbf{k}''}(p_\mathbf{k}^*p_{\mathbf{k}''}-
|p_{\mathbf{k}''}|^2)+p_\mathbf{k}^*\mathcal{E}_\text{p}d_\mathrm{cv}\Big]-\nonumber \\
-& 2V_{\mathbf{k}-\mathbf{k}'}(p_\mathbf{k}p_{\mathbf{k}'}^*-|p_\mathbf{k}|^2).
\end{align}
According to the first equation the amplitude $a_\mathbf{k}$ should be linear with $\mathcal{E}_\text{t}$, 
while $b_\mathbf{k}\sim 0$. 
Hence, the dominant contribution appears from $a_\mathbf{k}$, and we put all $b_\mathbf{k}=0$ in further calculations.  
Then the simplified equation reads 
\begin{align}
\label{eq:eff_eq_p_plus}
\sum_{\mathbf{k}'}\Big[H^0_{\mathbf{k}\mathbf{k}'}+\delta H_{\mathbf{k}\mathbf{k}'}-\hbar\omega_\text{t}
\delta_{\mathbf{k}\mathbf{k}'}\Big]a_{\mathbf{k}'}=
(1-2|p_\mathbf{k}|^2)d_\mathrm{cv}\mathcal{E}_\text{t}. 
\end{align}
It is convenient to introduce the substitution $a_\mathbf{k}=\sum_\lambda a_\lambda \psi_{\lambda,\mathbf{k}}$, 
where $\psi_{\lambda,\mathbf{k}}$ are eigenfunctions of the $H^0_{\mathbf{k}\mathbf{k}'}$ matrix with eigenvalues 
$\hbar\omega_\lambda$
\begin{equation}
\sum_{\mathbf{k}'} H^0_{\mathbf{k}\mathbf{k}'}\psi_{\lambda,\mathbf{k}'}=\hbar\omega_\lambda\psi_{\lambda,\mathbf{k}}.
\end{equation} 
The eigenfunctions $\psi_{\lambda,\mathbf{k}}$ are nothing but the exciton wave-functions in the $\mathbf{k}$-space
\begin{equation}
\sum_\mathbf{k} \psi^*_{\lambda,\mathbf{k}}\psi_{\lambda',\mathbf{k}}=\delta_{\lambda\lambda'}.
\end{equation}
Inserting the expansion 
$a_\mathbf{k}=\sum_\lambda a_\lambda \psi_{\lambda,\mathbf{k}}$ in the main equation, 
multiplying the result with $\psi_{\lambda,\mathbf{k}}^*$ and then summing over $\mathbf{k}$ yields  
\begin{align}
\sum_{\lambda'}\Big[(\hbar\omega_\lambda-\hbar\omega_\text{t})\delta_{\lambda\lambda'}+&
\delta H_{\lambda\lambda'}\Big]a_{\lambda'}=\nonumber \\ =&d_\mathrm{cv}\mathcal{E}_\text{t}\sum_\mathbf{k}
\psi_{\lambda,\mathbf{k}}^*(1-2|p_\mathbf{k}|^2). 
\end{align}
We decompose the matrix $\delta H_{\lambda\lambda'}$ into two parts
\begin{equation}
\delta H_{\lambda\lambda'}=\Pi_{\lambda\lambda'}+\Delta_{\lambda\lambda'}.
\end{equation}
The first term 
\begin{equation}
\Pi_{\lambda\lambda'}=2\mathcal{E}_\text{p}d_\mathrm{cv}\sum_\mathbf{k} \psi_{\lambda,\mathbf{k}}^*p_\mathbf{k}^*
\psi_{\lambda',\mathbf{k}}
\end{equation}
corresponds to the non-linear {\it exciton-pump-field interaction},
while the second term 
\begin{equation}
\Delta_{\lambda\lambda'}=2\sum_{\mathbf{k},\mathbf{k}'}V_{\mathbf{k}-\mathbf{k}'}\psi_{\lambda,\mathbf{k}}^*
(p_\mathbf{k}^*-p_{\mathbf{k}'}^*)(p_\mathbf{k}\psi_{\lambda',\mathbf{k}'}+
p_{\mathbf{k}'}\psi_{\lambda',\mathbf{k}}),
\end{equation}
describes the so-called {\it exciton-exciton interaction} (see details in \cite{Ell1989}). 
One can write the solution in the form 
\begin{align}
\label{eq:a_lambda}
a_{\lambda}=\frac{d_\mathrm{cv}\mathcal{E}_\text{t}\sum_\mathbf{k} \psi_{\lambda,\mathbf{k}}^*(1-2|p_\mathbf{k}|^2)-
\sum_{\lambda'\neq\lambda}\delta H_{\lambda,\lambda'}a_{\lambda'}}{\hbar{(\bar{\omega}_\lambda-\omega_\text{t}-i0)}},
\end{align}
where we have introduced the renormalized exciton energies
\begin{equation}
\bar{\omega}_\lambda=\omega_\lambda+\delta H_{\lambda\lambda}/\hbar.
\end{equation} 
The corresponding solution manifests the existence of the optical transitions at energies $\bar{\omega}_\lambda$, 
see \cite{Haug2009}. 
Hence, $\delta H_{\lambda\lambda}$ are nothing but the excitonic energy shifts in the presence of the 
non-resonant pump field. Therefore, we conclude that $\sigma_\text{p}^+/\sigma_\text{t}^+$ 
configuration of the pump and test beams induces the optical transitions in the K$^+$ point of 
monolayer and shifts the energy of the corresponding excitons.

Note that the denominator of $a_\lambda$ in Eq.~(\ref{eq:a_lambda}) 
contains an infinitesimally small imaginary part which implies a
zero broadening of the corresponding exciton line. It is a result of our approximation described before. 
The realistic broadening of the exciton line can be introduced phenomenologically by adding the dissipation term 
 $-i\gamma \delta p_\mathbf{k}$ on the r.h.s. of Eq.~(\ref{eq:delta_p}) and repeating all the steps of the derivation   of its solution.     

Taking into account the polarization induced by the pump field (see Appendix~\ref{app:polarization})  
\begin{equation}
p_\mathbf{k}\approx\mathcal{E}_\text{p}d_\mathrm{cv}\sqrt{S}
\frac{\psi_{1s,\mathbf{k}}\psi_{1s}(\mathbf{r}=0)}{E_{1s}-\hbar\omega_\text{p}}, 
\end{equation}
we obtain 
\begin{align}
\Pi_{\lambda\lambda}=&\frac{2|d_\mathrm{cv}|^2\mathcal{E}_\text{p}^2}{E_{1s}-\hbar\omega_\text{p}}\Big[\sqrt{S}\psi_{1s}(\mathbf{r}=0)\sum_\mathbf{k}
|\psi_{\lambda,\mathbf{k}}|^2 \psi_{1s,\mathbf{k}}\Big]=\nonumber \\=
&\frac{2|d_\mathrm{cv}|^2\mathcal{E}_\text{p}^2}{E_{1s}-\hbar\omega_\text{p}}\rho_\lambda.
\end{align}
The answer deviates form the standard Bloch shift of two-level system by an enhancement factor 
$\rho_\lambda$. 
For the case of the $1s$ exciton we have 
\begin{equation}
\Pi_{1s1s}=\frac{2|d_\mathrm{cv}|^2\mathcal{E}_\text{p}^2}{E_{1s}-\hbar\omega_\text{p}}\rho_{1s}.
\end{equation} 
Let us calculate the exciton-exciton interaction correction to the energy shift of the exciton
\begin{align}
\Delta_{\lambda\lambda}=&\frac{2|d_\mathrm{cv}|^2\mathcal{E}_\text{p}^2}{(E_{1s}-\hbar\omega_\text{p})^2}
S[\psi_{1s}(\mathbf{r}=0)]^2
\sum_{\mathbf{k},\mathbf{k}'}V_{\mathbf{k}-\mathbf{k}'}\psi_{\lambda,\mathbf{k}}^*\times \nonumber \\
\times&
(\psi_{1s,\mathbf{k}}-\psi_{1s,\mathbf{k}'})(\psi_{1s,\mathbf{k}}\psi_{\lambda,\mathbf{k}'}+
\psi_{1s,\mathbf{k}'}\psi_{\lambda,\mathbf{k}})=\nonumber \\ 
=&\frac{4|d_\mathrm{cv}|^2\mathcal{E}_\text{p}^2}{(E_{1s}-\hbar\omega_\text{p})^2}\eta_{\lambda},
\end{align}
which  for the 1s exciton case transforms into 
\begin{align}
\Delta_{1s1s}=\frac{4|d_\mathrm{cv}|^2\mathcal{E}_\text{p}^2}{(E_{1s}-\hbar\omega_\text{p})^2}\eta_{1s}.
\end{align}
The numerical values of $\eta_{1s}$ and $\rho_{1s}$ are estimated in Appendix~\ref{app:estimation}.

\subsection{$\sigma_\text{p}^+/\sigma_\text{t}^+$ case, $\tau=-1$}

We consider the processes in the K$^-$ point. 
Then, the dipole moment matrix element is $d^\tau_\mathrm{cv}=-d_\mathrm{cv}$,  
electric field of the pulses reads
\begin{equation}
\mathcal{E}=\mathcal{E}_\text{p}e^{i\omega_\text{p}t}+\mathcal{E}_\text{t}e^{i\omega_\text{t}t}.
\end{equation}    
The derivation of the Bloch equations of motion can be done analogous to how it was done before. Therefore, 
one obtains them by replacing $d_\mathrm{cv}\rightarrow -d_\mathrm{cv}$,
$\omega_\text{p}\rightarrow -\omega_\text{p}$, and $\omega_\text{t}\rightarrow -\omega_\text{t}$, 
\begin{align}
\sum_{\mathbf{k}'}\Big[&H^0_{\mathbf{k}\mathbf{k}'}+\delta H_{\mathbf{k}\mathbf{k}'}+
\hbar\omega_\text{t}\delta_{\mathbf{k}\mathbf{k}'}\Big]a_{\mathbf{k}'}=
(2|p_\mathbf{k}|^2-1)d_\mathrm{cv}\mathcal{E}_\text{t} + \nonumber \\&+
2\sum_{\mathbf{k}'}V_{\mathbf{k}-\mathbf{k}'}p_\mathbf{k}p_{\mathbf{k}'}(b_{\mathbf{k}'}^*-b_\mathbf{k}^*)+
2p_\mathbf{k}\mathcal{E}_\text{p}d_\mathrm{cv}b_\mathbf{k}^*,\\
\sum_{\mathbf{k}'}\Big[&H^0_{\mathbf{k}\mathbf{k}'}+\delta H_{\mathbf{k}\mathbf{k}'}-
\hbar(\omega_\text{t}-2\omega_\text{p})\delta_{\mathbf{k}\mathbf{k}'}\Big]b_{\mathbf{k}'}=\nonumber \\&=
2\sum_{\mathbf{k}'}V_{\mathbf{k}-\mathbf{k}'}p_\mathbf{k}p_{\mathbf{k}'}(a_{\mathbf{k}'}^*-a_\mathbf{k}^*)+
2p_\mathbf{k}\mathcal{E}_\text{p}d_\mathrm{cv}a_\mathbf{k}^*.
\end{align} 
Now we see that the first equation contains a non-resonant term on the l.h.s. and therefore the dominant solution 
for $a_\mathbf{k}$ doesn't allow optical transitions. 
The second equation contains a resonant term on the l.h.s., however it does't contain $\mathcal{E}_\text{t}$ terms on the r.h.s., and hence $b_\mathbf{k}$ coefficients don't give the leading contributions to the optical susceptibility of the monolayer in this case. We conclude that the $\sigma_\text{p}^+/\sigma_\text{t}^+$ configuration of the pump and test beams does not induce optical transitions in the K$^-$ point of the monolayer. 

\subsection{$\sigma_\text{p}^+/\sigma_\text{t}^-$ case, $\tau=1$}

We consider again the processes in the K$^+$ point. 
The dipole moment matrix element is $d^\tau_\mathrm{cv}=d_\mathrm{cv}$ and 
electric field of the pulses reads
\begin{equation}
\mathcal{E}=\mathcal{E}_\text{p}e^{-i\omega_\text{p}t}+\mathcal{E}_\text{t}e^{i\omega_\text{t}t}. 
\end{equation}
Repeating the same steps of the derivation and keeping the same definitions introduced in the previous subsection A we get the following system of equations for the $a_k$ and $b_k$ coefficients. This set of equations can be derived from equations for the $\sigma_\text{p}^+/\sigma_\text{t}^+$, $\tau=1$ case by replacing 
$\omega_\text{t}\rightarrow -\omega_\text{t}$
\begin{align}
\sum_{\mathbf{k}'}\Big[&H^0_{\mathbf{k}\mathbf{k}'}+\delta H_{\mathbf{k}\mathbf{k}'}+
\hbar\omega_\text{t}\delta_{\mathbf{k}\mathbf{k}'}\Big]a_{\mathbf{k}'}=(1-2|p_\mathbf{k}|^2)
d_\mathrm{cv}\mathcal{E}_\text{t} + \nonumber \\&+ 2\sum_{\mathbf{k}'}V_{\mathbf{k}-\mathbf{k}'}
p_\mathbf{k}p_{\mathbf{k}'}
(b_{\mathbf{k}'}^*-b_\mathbf{k}^*)-2p_\mathbf{k}\mathcal{E}_\text{p}d_\mathrm{cv}b_\mathbf{k}^*,\\
\sum_{\mathbf{k}'}\Big[&H^0_{\mathbf{k}\mathbf{k}'}+\delta H_{\mathbf{k}\mathbf{k}'}-
\hbar(\omega_\text{t}+2\omega_\text{p})\delta_{\mathbf{k}\mathbf{k}'}\Big]b_{\mathbf{k}'}=\nonumber \\&=
2\sum_{\mathbf{k}'}V_{\mathbf{k}-\mathbf{k}'}p_\mathbf{k}p_{\mathbf{k}'}(a_{\mathbf{k}'}^*-a_\mathbf{k}^*)
-2p_\mathbf{k}\mathcal{E}_\text{p}d_\mathrm{cv}a_\mathbf{k}^*.
\end{align}
As one can see the first equation for dominant component $a_\mathbf{k}$ does not contain a resonant term, 
and hence it does not lead to exciton transitions in the K$^+$ point. The resonant term exists in the 
second equation, however this term is not leading. Hence, the $\sigma^-$ polarized test beam does not 
induce the exciton transitions in the K$^+$ point.      

\subsection{$\sigma_\text{p}^+/\sigma_\text{t}^-$ case, $\tau=-1$}
\label{subsec:tau_minus}
We consider the processes in the K$^-$ point. 
The dipole moment matrix element is $d^\tau_\mathrm{cv}=-d_\mathrm{cv}$, and  
electric field of the pulses reads
\begin{equation}
\mathcal{E}=\mathcal{E}_\text{p}e^{i\omega_\text{p}t}+\mathcal{E}_\text{t}e^{-i\omega_\text{t}t}.
\end{equation}    
The derivation of the equations of the motion can be done from the equations for the
$\sigma_\text{p}^+/\sigma_\text{t}^+$, $\tau=1$ case by replacing $d_\mathrm{cv}\rightarrow -d_\mathrm{cv}$, 
$\omega_\text{p}\rightarrow -\omega_\text{p}$
\begin{align}
\sum_{\mathbf{k}'}\Big[&H^0_{\mathbf{k}\mathbf{k}'}+\delta H_{\mathbf{k}\mathbf{k}'}-
\hbar\omega_\text{t}\delta_{\mathbf{k}\mathbf{k}'}\Big]a_{\mathbf{k}'}=
(2|p_\mathbf{k}|^2-1)d_\mathrm{cv}\mathcal{E}_\text{t} + \nonumber \\&+
2\sum_{\mathbf{k}'}V_{\mathbf{k}-\mathbf{k}'}p_\mathbf{k}p_{\mathbf{k}'}(b_{\mathbf{k}'}^*-b_\mathbf{k}^*)
+2p_\mathbf{k}\mathcal{E}_\text{p}d_\mathrm{cv}b_\mathbf{k}^*,\\
\sum_{\mathbf{k}'}\Big[&H^0_{\mathbf{k}\mathbf{k}'}+\delta H_{\mathbf{k}\mathbf{k}'}+
\hbar(\omega_\text{t}+2\omega_\text{p})\delta_{\mathbf{k}\mathbf{k}'}\Big]b_{\mathbf{k}'}=\nonumber \\&=
2\sum_{\mathbf{k}'}V_{\mathbf{k}-\mathbf{k}'}p_\mathbf{k}p_{\mathbf{k}'}(a_{\mathbf{k}'}^*-a_\mathbf{k}^*)+
2p_\mathbf{k}\mathcal{E}_\text{p}d_\mathrm{cv}a_\mathbf{k}^*.
\end{align} 
The first equation contains a resonant term, and hence the $\sigma^-$ polarized test beam induces the exciton transitions in K$^-$ point. For the case of $\tau=-1$ we have
\begin{equation}
H^0_{\mathbf{k}\mathbf{k}'}=\Big(\widetilde{E}_\mathrm{g}+\frac{\hbar^2k^2}{2m}\Big)\delta_{\mathbf{k}\mathbf{k}'}-
V_{\mathbf{k}-\mathbf{k}'},
\end{equation}
\begin{align}
\label{eq:hkk_minus}
\delta H_{\mathbf{k}\mathbf{k}'}=&2\delta_{\mathbf{k}\mathbf{k}'}
\Big[\sum_{\mathbf{k}''}V_{\mathbf{k}-\mathbf{k}''}(p_\mathbf{k}^*p_{\mathbf{k}''}-|p_{\mathbf{k}''}|^2)-
p_\mathbf{k}^*\mathcal{E}_\text{p}d_\mathrm{cv}\Big]-\nonumber \\
-&2V_{\mathbf{k}-\mathbf{k}'}(p_\mathbf{k}p_{\mathbf{k}'}^*-|p_\mathbf{k}|^2).
\end{align}
According to the first equation the amplitude $a_\mathbf{k}$ should be linear with 
$\mathcal{E}_\text{t}$, while $b_\mathbf{k}\sim 0$. Hence, the dominant contribution appears 
from $a_\mathbf{k}$, and we put all $b_\mathbf{k}=0$ in further calculations.  
Then the simplified equation is 
\begin{align}
\label{eq:eff_eq_p_minus}
\sum_{k'}\Big[H^0_{\mathbf{k}\mathbf{k}'}+\delta H_{\mathbf{k}\mathbf{k}'}-\hbar\omega_\text{t}\delta_{\mathbf{k}\mathbf{k}'}\Big]a_{\mathbf{k}'}=(2|p_\mathbf{k}|^2-1)d_\mathrm{cv}\mathcal{E}_\text{t}. 
\end{align}
Introducing the substitution $a_\mathbf{k}=\sum_\lambda a_\lambda \psi_{\lambda,\mathbf{k}}$ and repeating the calculations done in the previous Sec.~\ref{subsec:tau_plus} we obtain  
\begin{align}
\sum_{\lambda'}\Big[(\hbar\omega_\lambda-\hbar\omega_\text{t})\delta_{\lambda\lambda'}+&\delta H_{\lambda\lambda'}\Big]
a_{\lambda'}=\nonumber \\=&
d_\mathrm{cv}\mathcal{E}_\text{t}\sum_\mathbf{k} \psi_{\lambda,\mathbf{k}}^*(2|p_\mathbf{k}|^2-1). 
\end{align}
We decompose the  matrix $\delta H_{\lambda\lambda'}$ into two parts
\begin{equation}
\delta H_{\lambda\lambda'}=\Pi_{\lambda\lambda'}+\Delta_{\lambda\lambda'}.
\end{equation}
The term 
\begin{equation}
\Pi_{\lambda\lambda'}=-2\mathcal{E}_\text{p}d_\mathrm{cv}\sum_\mathbf{k} \psi_{\lambda,\mathbf{k}}^*p_\mathbf{k}^*
\psi_{\lambda',\mathbf{k}}
\end{equation}
corresponds to the non-linear interaction between the exciton and the pump field,
while the second term 
\begin{equation}
\Delta_{\lambda\lambda'}=2\sum_{\mathbf{k},\mathbf{k}'}V_{\mathbf{k}-\mathbf{k}'}\psi_{\lambda,\mathbf{k}}^*
(p_\mathbf{k}^*-p_{\mathbf{k}'}^*)(p_\mathbf{k}\psi_{\lambda',\mathbf{k}'}+p_{\mathbf{k}'}\psi_{\lambda',\mathbf{k}}),
\end{equation}
describes the exciton-exciton interaction. 
One can write the solution in the form 
\begin{align}
a_{\lambda}=\frac{-d_\mathrm{cv}\mathcal{E}_\text{t}\sum_\mathbf{k} \psi_{\lambda,\mathbf{k}}^*(1-2|p_\mathbf{k}|^2)-
\sum_{\lambda'\neq\lambda}\delta H_{\lambda\lambda'}a_{\lambda'}}{\hbar{(\bar{\omega}_\lambda-\omega_\text{t}-i0)}}, 
\end{align}
where we have introduced the renormalized exciton energies
\begin{equation}
\bar{\omega}_\lambda=\bar{\omega}_\lambda+\delta H_{\lambda\lambda}/\hbar.
\end{equation} 
Therefore, $\delta H_{\lambda\lambda}$ are again the excitonic energy shifts in the presence of the 
non-resonant pump field. This result is similar to the $\sigma_\text{p}^+/\sigma_\text{t}^+$, $\tau=1$ case. Therefore we can repeat 
the same steps of calculations from the previous Sec.~\ref{subsec:tau_plus} to obtain the excitonic energy shifts. 
The polarization $p_\mathbf{k}$ induced by the pump field is
\begin{equation}
p_\mathbf{k}\approx-\mathcal{E}_\text{p}d_\mathrm{cv}\sqrt{S}
\frac{\psi_{1s,\mathbf{k}}\psi_{1s}(\mathbf{r}=0)}{E_{1s}+\hbar\omega_\text{p}}. 
\end{equation}  
Then one gets 
\begin{align}
\Pi_{\lambda\lambda}=&
\frac{2|d_\mathrm{cv}|^2\mathcal{E}_\text{p}^2}{E_{1s}+\hbar\omega_\text{p}}\rho_\lambda, \\
\Delta_{\lambda\lambda}=&
\frac{4|d_\mathrm{cv}|^2\mathcal{E}_\text{p}^2}{(E_{1s}+\hbar\omega_\text{p})^2}\eta_{\lambda},
\end{align}
This result coincides with the result of the $\sigma_\text{p}^+/\sigma_\text{t}^+$, $\tau=1$ case, except the sign before $\hbar\omega_\text{p}$ in the denominator.  
 
The expressions for $\rho_{\lambda}$ and $\eta_{\lambda}$ are the same as the ones in the Sec.~\ref{subsec:tau_plus}, 
their numerical values for the particular case of 1s exciton are calculated in the 
Appendix~\ref{app:estimation}.
 
\section{Estimate of the energy shifts for WS\lowercase{e}$_2$}
\label{sec:Estimate_of_the_energy_shifts_for_WSe2}

The effective dielectric constant of the studied system lies in between the dielectric constants of the Si/SiO$_2$ substrate and the dielectric constant of the hBN flake. This is an important parameter since it modifies the binding energy of the excitons $E_\text{b}$, the bandgap $\widetilde{E}_\mathrm{g}$ in the system and hence the energy of 1s  excitonic state  $E_{1s}=\widetilde{E}_\mathrm{g}-E_\text{b}$.

Let us consider two limit cases. For the case of the Si/SiO$_2$ substrate we have $\widetilde{E}_\mathrm{g}= 2.02\,\mathrm{eV}$, the energy of 
1s A-exciton $E_A=1.639\,\mathrm{eV}$ and 
$E_\text{b}=0.37\,\mathrm{eV}$ \cite{He2014}. 
For the case of the WSe$_2$ flake encapsulated in hBN we have $\widetilde{E}_\mathrm{g}=1.873\,\mathrm{eV}$ and 
$E_A=1.706\,\mathrm{eV}$, $E_\text{b}=0.167\,\mathrm{eV}$ \cite{Molas2019}. 
The experimental value $E_A=1.639\,\mathrm{eV}$ surprisingly coincides with the first case. Therefore we should suppose that the considered sample is not screened effectively by the top hBN flake. Thus, we consider the parameters of the first case as the source for our further calculations. 

First, using the values of the binding energy and reduced exciton mass $m=0.21m_0$ ($m_0$ is the bare electron mass) we estimate the effective dielectric constant $\varepsilon$, with the help of the variational method from 
Ref.~[\onlinecite{Molas2019}]. It gives us $\varepsilon\approx 1.6$ and the coefficient 
$\beta r_0/\varepsilon \approx 4.564$ ($r_0$ is the screening length for Rytova-Keldysh potential) for the trial 1s exciton wave-function $\psi_0(\mathbf{r})=\beta\exp(-\beta r/2)/\sqrt{2\pi}$. 

To estimate the energy shifts of 1s A-exciton we use the formula (see Appendix~\ref{app:estimation} for details)
\begin{equation}
\Delta E_\pm^A=\frac{2|d_\mathrm{cv}|^2\mathcal{E}_\text{p}^2}{E_A\mp\hbar\omega_\text{p}}
\Big[\frac{16}{7}+\frac{2\eta_{1s}}{E_A\mp\hbar\omega_\text{p}}\Big].
\end{equation}
Using the definition of $|d_\text{cv}|$ (see Appendix~\ref{app:two_band_model}) we
present the Rabi shift in the form    
\begin{equation}
E_{R,\pm}^A=\frac{2|d_\mathrm{cv}|^2\mathcal{E}_\text{p}^2}{E_A\mp\hbar\omega_\text{p}}\approx
\frac{2v^2e^2\mathcal{E}^2_\text{p}}{\bar{E}_\mathrm{g}^2(E_A\mp\hbar\omega_\text{p})},
\end{equation}
where $\bar{E}_\text{g}$ and $v$ are the single-particle band gap and Dirac velocity of the monolayer,
$e$ is the elementary electron charge.

There is some uncertainty in the precise values of $\bar{E}_\text{g}$ parameter:  
$\bar{E}_\text{g}\approx 1.337\,\text{eV}$ (derived from Ref.~[\onlinecite{Kormanyos2015}]) and 
$\bar{E}_\text{g}\approx 1.435\,\text{eV}$ (derived from Refs.~[\onlinecite{Fang2015},\onlinecite{Fang2018}]). 
We use an average value $\bar{E}_\text{g}\approx 1.386\,\text{eV}$
in further calculations.
Then, taking $v\approx 3.382 \mathrm{eV}\cdot\mbox{\AA}$ 
(see details in Ref.~[\onlinecite{Fang2018}]), 
$\hbar\omega_\text{p}=0.62\,\mathrm{eV}$, $E_A\approx 1.639\,\mathrm{eV}$ 
(which corresponds to wavelength $\lambda_A=750\,\mathrm{nm}$)
we get the following expressions
\begin{align}
E_{R,+}^A/\mathcal{E}_\text{p}^2 \approx 11.7\,\,\mathrm{eV}\!\cdot\!\mbox{\AA}^2/\text{V}^2,\\ 
E_{R,-}^A/\mathcal{E}_\text{p}^2 \approx 5.3\,\,\mathrm{eV}\!\cdot\!\mbox{\AA}^2/\text{V}^2.
\end{align}
Then evaluating the parameter $\eta_{1s}$ for $a=\beta r_0/2\varepsilon=2.282$
\begin{equation}
\eta_{1s}=\frac{4\beta e^2}{\varepsilon\pi^2}I(a)=\frac{8e^2}{\pi^2r_0}aI(a)\approx 298\,\textrm{meV},
\end{equation}
we obtain
\begin{align}
\Delta E^A_+/\mathcal{E}_\text{p}^2 \approx  33.6\,\,\mathrm{eV}\!\cdot\!\mbox{\AA}^2/\text{V}^2,\\
\Delta E^A_-/\mathcal{E}_\text{p}^2 \approx  13.5\,\,\mathrm{eV}\!\cdot\!\mbox{\AA}^2/\text{V}^2.
\end{align}
Note that, the exciton-exciton interaction contribution gives around $20\%$ and $10\%$ 
of the full exciton shifts in the K$^+$ and K$^-$ points, respectively.  
The ratio of OS and BS energy shifts is $\Delta E_+^A/\Delta E_-^A\approx 2.48$.

The previous strategy can be applied for the calculation of the energy shifts of the B-excitons.   
\begin{equation}
\Delta E_\pm^B=\frac{2|d_\mathrm{cv}|^2\mathcal{E}_\text{p}^2}{E_B\mp\hbar\omega_\text{p}}
\Big[\frac{16}{7}+\frac{2\eta_{1s}}{E_B\mp\hbar\omega_\text{p}}\Big],
\end{equation}
where the parameters $d_\mathrm{cv}$, $\eta_{1s}$ should be recalculated. 
The exciton energy is $E_B=2.083\,\textrm{eV}$. To estimate $|d_\mathrm{cv}|\approx |e|v/\bar{E}_\mathrm{g}$ 
one needs to know the band-gap $\bar{E}_\mathrm{g}$ between the pair of the bands for B-excitons. 
According to Refs.[\onlinecite{Kormanyos2015},\onlinecite{Fang2015}] it is $\bar{E}_\mathrm{g}=1.766$ and 
$\bar{E}_\mathrm{g}=1.866\,\mathrm{eV}$, respectively.
We take the average value $\bar{E}_\mathrm{g}=1.816\,\mathrm{eV}$ for further calculations.  
Using the reduced mass of B-exciton $m=0.23m_0$ (see \cite{Kormanyos2015}) and $\varepsilon=1.6$
we get $E_\text{b}=0.381\,\mathrm{eV}$. 
The obtained binding energy corresponds to $\beta r_0/\varepsilon\approx 4.804$.
Then the Rabi shifts  
\begin{equation}
E_{R,\pm}^B\approx \frac{2v^2 e^2\mathcal{E}_\text{p}^2}{\bar{E}_\mathrm{g}^2(E_B\mp\hbar\omega_\text{p})},
\end{equation}
for the case $v\approx 3.382\,\mathrm{eV}\cdot\mbox{\AA}$, 
$\bar{E}_\mathrm{g}=1.816\,\mathrm{eV}$,
$\hbar\omega_\text{p}=0.62\,\mathrm{eV}$, $E_B\approx 2.083\,\mathrm{eV}$ 
satisfy the following equalities 
\begin{align}
E_{R,+}^B/\mathcal{E}_\text{p}^2 \approx 4.7\,\,\mathrm{eV}\!\cdot\!\mbox{\AA}^2/\text{V}^2,\\ 
E_{R,-}^B/\mathcal{E}_\text{p}^2 \approx  2.6\,\,\mathrm{eV}\!\cdot\!\mbox{\AA}^2/\text{V}^2.
\end{align}  
The parameter $\eta_{1s}$ for $a=\beta r_0/2\varepsilon=2.402$ is
\begin{equation}
\eta_{1s}=\frac{8e^2}{\pi^2r_0}aI(a)\approx 302\,\textrm{meV}.
\end{equation}
Then we have
\begin{align}
\Delta E_+^B/\mathcal{E}_\text{p}^2 \approx  12.7\,\,\mathrm{eV}\!\cdot\!\mbox{\AA}^2/\text{V}^2, \\
\Delta E_-^B/\mathcal{E}_\text{p}^2 \approx  6.5\,\,\mathrm{eV}\!\cdot\!\mbox{\AA}^2/\text{V}^2.
\end{align}
The exciton-exciton interaction contribution is around $15\%$ and $9\%$ 
of the total energy shift in the K$^+$ and K$^-$ points, respectively.  
The ratio of the OS and BS energy shifts is 
$\Delta E^B_+/\Delta E^B_-\approx 1.95$.

To conclude we have calculated the corresponding OS and BS shifts in WSe$_2$ monolayer for
the experimental value of intensity of the pump field and compared the obtained results with the experimental ones.
They are presented in Tab.~\ref{tab:tab_1}

\begin{table}[h]
 \begin{center}
 \begin{tabular}{ccccc}
 \hline\hline \\ [-0.5ex]
 { $\Delta E$\, [meV] } & OS,1sA & BS,1sA & OS,1sB & BS,1sB  \\ [1.0ex]
 \hline \\
 Theor.  & $38$ & $15$ & $14$ & $7$ \\ [1.2ex]
 Exp.    & $23.4\pm 0.7$ & $14.4\pm0.5$  & $3.6\pm0.4$ & $2.3\pm0.4$ \\ [1.2ex]
 \hline
\end{tabular}
\caption{OS ($\Delta E_\text{OS}$) and BS ($\Delta E_\text{BS}$) (in meV) shifts for 1sA and 1sB excitons in WSe$_2$ 
    monolayer, calculated theoretically (Theor.) and compared with the corresponding experimental values (Exp.) 
		at pump intensity $\mathcal{E}=30\,\mathrm{GW/cm^2}$. }
		\label{tab:tab_1}
\end{center}
\end{table}	
For the case of A excitonic transitions the theoretical and experimental results are quite similar. 
The strong deviation of the theoretical estimates and experimental results for B excitonic transitions
can indicate a smaller Fermi velocity $v$ and/or larger single-particle band gap $\bar{E}_\text{g}$ than 
those used in the current study.   

\section{Estimate of the energy shifts for M\lowercase{o}S$_2$}
\label{sec:Estimate_of_the_energy_shifts_for_MoS2}

Since all TMD samples were prepared with the same approach 
we suppose that the dielectric constant of the surrounding medium 
for MoS$_2$ is the same as in the previous case $\varepsilon=1.6$.  
To estimate the energy shift of 1s A-excitons we use the formula 
\begin{equation}
\Delta E_\pm^A=\frac{2|d_\mathrm{cv}|^2\mathcal{E}_\text{p}^2}{E_A\mp\hbar\omega_\text{p}}
\Big[\frac{16}{7}+\frac{2\eta_{1s}}{E_A\mp\hbar\omega_\text{p}}\Big].
\end{equation}
The Rabi shift  
\begin{equation}
E_{R,\pm}^A\approx
\frac{2v^2 e^2\mathcal{E}_\text{p}^2}{\bar{E}_\mathrm{g}^2(E_A\mp\hbar\omega_\text{p})},
\end{equation}
for the case $v=3.373\,\mathrm{eV}\cdot\mbox{\AA}$ 
(see details in \cite{Fang2018}), $\bar{E}_\mathrm{g}\approx 1.69\,\mathrm{eV}$ 
(as in the previous case we take an average value of the band gap energies
$\bar{E}_\mathrm{g}\approx 1.67\,\mathrm{eV}$ from \cite{Kormanyos2015} 
and $\bar{E}_\mathrm{g}\approx 1.71\,\mathrm{eV}$ from \cite{Fang2018}),
$\hbar\omega_\text{p}=0.62\,\mathrm{eV}$,
$E_A\approx 1.886\,\mathrm{eV}$, we get the following expressions
\begin{align}
E_{R,+}^A/\mathcal{E}_\text{p}^2 \approx 6.3\,\,\mathrm{eV}\!\cdot\!\mbox{\AA}^2/\text{V}^2,\\ 
E_{R,-}^A/\mathcal{E}_\text{p}^2 \approx 3.2\,\,\mathrm{eV}\!\cdot\!\mbox{\AA}^2/\text{V}^2.
\end{align}
Then using the dielectric constant $\varepsilon=1.6$, 
reduced mass $m=0.26m_0$ (see \cite{Molas2019,Berkelbach2013,Goryca2019}) and 
$r_0\approx 41.5\,\mathrm{\AA}$ (see \cite{Berkelbach2013}) we obtain the binding energy 
$E_\text{b}\approx 418\,\text{meV}$ and the following value for $\beta r_0/\varepsilon\approx 4.917$ 
for the trial wave-function of the 1s exciton. 
The parameter $\eta_{1s}$ for $a=\beta r_0/2\varepsilon\approx 2.459$ is
\begin{equation}
\eta_{1s}=\frac{8e^2}{\pi^2r_0}aI(a)\approx 329.7\,\textrm{meV}.
\end{equation}
Then we obtain 
\begin{align}
\Delta E_+^A/\mathcal{E}_\text{p}^2 \approx 17.7\,\,\mathrm{eV}\!\cdot\!\mbox{\AA}^2/\text{V}^2, \\
\Delta E_-^A/\mathcal{E}_\text{p}^2 \approx 8.1\,\,\mathrm{eV}\!\cdot\!\mbox{\AA}^2/\text{V}^2.
\end{align}
The exciton-exciton interaction contribution is around $19\%$ and $10\%$ 
of the total energy shift in the K$^+$ and K$^-$ points, respectively.  
The ratio of the OS and BS energy shifts is
$\Delta E_+^A/\Delta E_-^A\approx 2.19$.

To calculate the energy shifts of the B-excitons we use the formula
\begin{equation}
\Delta E_\pm^B=\frac{2|d_\mathrm{cv}|^2\mathcal{E}_\text{p}^2}{E_B\mp\hbar\omega_\text{p}}
\Big[\frac{16}{7}+\frac{2\eta_{1s}}{E_B\mp\hbar\omega_\text{p}}\Big],
\end{equation}
with redefined parameters $d_\mathrm{cv}$, $\eta_{1s}$. 
The exciton energy is $E_B=2.032\,\textrm{eV}$. To estimate $|d_\mathrm{cv}|\approx |e|v/E_\mathrm{g}$ 
we use the band-gap $\bar{E}_\mathrm{g}\approx 1.844$, which is an average value of  
$\bar{E}_\mathrm{g}\approx 1.821$ from \cite{Kormanyos2015} and $\bar{E}_\mathrm{g}\approx1.866\,\text{eV}$ 
from for \cite{Fang2018}.  
Then the Rabi shifts
\begin{align}
E_{R,\pm}^B\approx \frac{2v^2 e^2\mathcal{E}_\text{p}^2}{\bar{E}_\mathrm{g}^2(E_B\mp\hbar\omega_p)},
\end{align}
satisfy the equalities
\begin{align}
E_{R,+}^B/\mathcal{E}_\text{p}^2 \approx 4.7\,\,\mathrm{eV}\!\cdot\!\mbox{\AA}^2/\text{V}^2,\\ 
E_{R,-}^B/\mathcal{E}_\text{p}^2 \approx 2.5\,\,\mathrm{eV}\!\cdot\!\mbox{\AA}^2/\text{V}^2.
\end{align} 
The reduced mass $m=0.26m_0$ (see \cite{Kormanyos2015}) of B-exciton coincides with 
the reduced mass of A-exciton. Therefore, the parameters $\beta r_0/\varepsilon$ 
and  $\eta_{1s}$ are the same as in the previous case and we obtain
\begin{align}
\Delta E_+/\mathcal{E}_\text{p}^2 \approx 13.1\,\,\mathrm{eV}\!\cdot\!\mbox{\AA}^2/\text{V}^2, \\
\Delta E_-/\mathcal{E}_\text{p}^2 \approx 6.4\,\,\mathrm{eV}\!\cdot\!\mbox{\AA}^2/\text{V}^2.
\end{align}
The exciton-exciton interaction contribution is around $17\%$ and $10\%$ 
of the total energy shifts in the K$^+$ and K$^-$ points, respectively.  
The ratio of OS and BS energy shifts is $\Delta E_+/\Delta E_-\approx 2$.

To conclude we have calculated the corresponding the OS and BS shifts in MoS$_2$ monolayer for
the experimental value of intensities of the pump field. They are presented in Tab.~\ref{tab:tab_2}.
\begin{table}[h]
 \begin{center}
 \begin{tabular}{ccccc}
 \hline\hline \\ [-0.5ex]
 { $\Delta E$\, [meV] } & OS,1sA & BS,1sA & OS,1sB & BS,1sB  \\ [1.0ex]
 \hline \\
 Theor.  & $24$ & $11$  & $18$ & $9$ \\ [1.2ex]
 Exp.  & $14.2\pm0.6$ & $8.5\pm0.4$ & $13.0\pm0.5$ & $7.4\pm0.4$  \\ [1.2ex]
 \hline
\end{tabular}
\caption{OS ($\Delta E_\text{OS}$) and BS ($\Delta E_\text{BS}$) (in meV) shifts for 1sA and 1sB excitons in MoS$_2$
    monolayer, calculated theoretically (Theor.) and compared with the corresponding experimental values (Exp.)
		at pump intensity $\mathcal{E}=36\,\mathrm{GW/cm^2}$.}
		\label{tab:tab_2}
\end{center}
\end{table}	
One can observe that for the case of A and B excitonic transitions the theoretical and experimental 
results are quite similar. 

\section{Comparison with the previously obtained results}
\label{sec:other_results}

In order to compare our results with the previously obtained ones (see
Refs.~[\onlinecite{Kim2014},\onlinecite{Sie2015},\onlinecite{Sie2017},\onlinecite{Cunningham2019},
\onlinecite{LaMountain2018}]) we use the parameter $C$ 
introduced in Ref.~[\onlinecite{LaMountain2018}] for OS shift $\Delta E\equiv\Delta E_+^A$ of 1sA-exciton, with 
the energy detuning $(E_{A}-\hbar\omega_\text{p})$ and intensity of the applied pump pulse $I_\text{p}$
\begin{equation}
\label{eq:C}
C\equiv\Delta E\left(\frac{E_{A}-\hbar\omega_\text{p}}{I_\text{p}}\right).
\end{equation} 
This parameter cancels the evident intensity and energy dependences of the shifts obtained under different conditions, and hence it is a good observable to compare the results of different experiments. 
We compare only the results for the OS shifts of the 1sA exciton transitions, 
since the 1sB transitions and/or BS shifts are not represented widely in the literature. 
The results are summarized in Table~\ref{tab:tab_3}. 
\begin{table}[h]
 \begin{center}
 \begin{tabular}{ccccc}
 \hline\hline \\ [-0.5ex]
 { Material} & Substrate & Ref. & Method & $C$[eV$^2$cm$^2$/GW]   \\ [1.0ex]
 \hline \\
 WSe$_2$  & Sapphire &   [\onlinecite{Kim2014}]  & Exp.& $1.5\times 10^{-4}$  \\ [1.2ex]
 WS$_2$   & Sapphire &   [\onlinecite{Sie2015}]  & Exp.&  $9.2\times 10^{-3}$  \\  [1.2ex]
 WS$_2$   & Sapphire &   [\onlinecite{Sie2017}]  & Exp.& $8.7\times 10^{-4}$  \\  [1.2ex]
 WS$_2$   & Si/SiO$_2$ & [\onlinecite{Cunningham2019}] & Exp. &  $2.5\times 10^{-3}$  \\ [1.2ex]
 WSe$_2$  & Si/SiO$_2$ & [\onlinecite{LaMountain2018}] & Exp. &$5.3\times 10^{-5}$  \\ [1.2ex]
 MoS$_2$  & Si/SiO$_2$ & [\onlinecite{LaMountain2018}] & Exp. &$9.4\times 10^{-6}$  \\ [1.2ex]
 \hline \\ [-0.5ex]
 WSe$_2$  & Si/SiO$_2$ &  & Exp. &$ 8\times 10^{-4}$  \\ [1.2ex]
 MoS$_2$  & Si/SiO$_2$ &  & Exp. &$ 5\times 10^{-4}$  \\ [1.2ex]
 WSe$_2$  & Si/SiO$_2$ &  & Theor. &$1.3\times 10^{-3}$  \\ [1.2ex]
 MoS$_2$  & Si/SiO$_2$ &  & Theor. &$8.4\times 10^{-4}$  \\ [1.2ex]
 WS$_2$   & Sapphire &   [\onlinecite{Sie2015}]  & Theor.&  $1.3\times 10^{-3}$  \\  [1.2ex]
 \hline
\end{tabular}
\caption{Parameter $C$ [Eq.~(\ref{eq:C})] for the OS shifts of the 1sA exciton in TMD monolayers placed on different substrates. 
The values above the horizontal line are estimated from scientific literature, while the numbers below the horizontal line correspond to the findings of the current study. The labels (Theor.) and (Exp.) mark the method of obtaining of the corresponding values.}
		\label{tab:tab_3}
\end{center}
\end{table}	 
  
We calculated the corresponding values using the relation between the electric field 
$\mathcal{E}_\text{p}$ of the circularly polarized plane wave and its intensity 
$I_\text{p}=(c/4\pi)\mathcal{E}_\text{p}^2$ (in cgs units)
\begin{equation}
\frac{\mathcal{E}^2_\text{p}}{I_\text{p}}\approx 3.767\times 10^{-5}
\frac{\text{V}^2}{\mbox{\AA}^2}\frac{\text{cm}^2}{\text{GW}}, 
\end{equation} 
and the methodology, presented below for each paper separately. 

\begin{itemize}
	\item The authors of Ref.~[\onlinecite{Kim2014}] utilized the experimental formula 
$\Delta E=2S\mathcal{E}_\text{p}^2/(E_A-\hbar\omega_\text{p})$, with $S\approx 45\,\text{Debye}^2$, 
where $1\,\text{Debye}=0.2081943\, |e|\mbox{\AA}$. 
Therefore, for this case 
\begin{equation}
C_{[34]}=2S\frac{\mathcal{E}^2_\text{p}}{I_\text{p}}
\approx 1.469\times 10^{-4} \frac{\text{eV}^2\text{cm}^2}{\text{GW}}. 
\end{equation}

\item The authors of Ref.~[\onlinecite{Sie2015}] have the largest shift $\Delta E=18\,\text{meV}$, 
for the non-resonant pulse with energy detuning $E_A-\hbar\omega_\text{p}=180\,\text{meV}$, 
total flux $\mathcal{F}=120\,\mu\text{J}/\text{cm}^2$ and pulse duration (FWHM) $T=160\,\text{fs}$.  
Approximating the pulse profile by the Gaussian function $I(t)=I_\text{p}\exp(-\log(2)t^2/T^2)$ and
integrating it over the time we obtain 
\begin{equation}
\mathcal{F}=\int_{-\infty}^\infty dt I(t)=I_\text{p}T\sqrt{\frac{\pi}{\log(2)}},
\end{equation} 
which gives us $I_\text{p}\approx 0.352\,\text{GW}/\text{cm}^2$. Substituting this value into the formula for $C$, 
we obtain the following value $C_{[35]}\approx 9.197\times 10^{-3} \text{eV}^2\text{cm}^2/\text{GW}$.

  \item In Ref.~[\onlinecite{Sie2017}] the authors used the formula
\begin{equation}
\Delta E=\frac{\mu^2}{2}\frac{\mathcal{E}^2_\text{p}}{E_A-\hbar\omega}, 
\end{equation}
for the energy shift (see Eq.~(5) in the corresponding paper) and    
extracted $\mu=55\,\text{Debye}$ from their experimental results. 
Using that value for $\mu$ we obtain  
\begin{equation}
C_{[36]}=\frac{\mu^2}{2}\frac{\mathcal{E}_\text{p}^2}{I_\text{p}}\approx 2.47\times 10^{-3} 
\frac{\text{eV}^2\text{cm}^2}{\text{GW}}.
\end{equation}

 \item The authors of Ref.~[\onlinecite{Cunningham2019}] introduced an improved model to describe resonant OS shifts. 
We use the values $M_{gx}=5.2\,\text{Debye}$, $E_\text{b}=320\,\text{meV}$ and detuning energy 
$E_A-\hbar\omega_\text{p}=-23\,\text{meV}$ (see details in the description of Fig.~5e in the corresponding paper) 
to estimate $\Delta E$ for these parameters. Then applying the general formula for $C$ we obtain 
$C_{[37]}\approx 2.5\times 10^{-3}\text{eV}^2\text{cm}^2/\text{GW}$.

  \item Finally, in order to finish our considerations, we take the parameters of the energy $E_A=2$\,eV 
and $\hbar\omega_p=1.82$\,eV for WS$_2$ from Ref.~[\onlinecite{Sie2015}] and estimate the parameter $C$ using
our theoretical method. To achieve it we use the following parameters: 
$v\approx 3.882$\, eV$\cdot\mbox{\AA}$ [\onlinecite{Fang2018}], 
$\bar{E}_g\approx 1.9$\,eV [\onlinecite{Kormanyos2015},\onlinecite{Fang2018}],  dielectric constant of sapphire substrate $\varepsilon_{sapph}\approx 10$ [\onlinecite{Harman1993}], 
reduced exciton mass $\mu=0.15m_0$ [\onlinecite{Kormanyos2015},\onlinecite{Molas2019}], 
where $m_0$ is the free electron mass, and $r_0=37.89\,\mbox{\AA}$ [\onlinecite{Berkelbach2013}]. 
The obtained value of the parameter $C_\text{WS$_2$}^\text{theor.}\approx 1.3\times10^{-3}$\,eV$^2$cm$^2$/GW is in between the experimentally obtained values of Refs.~[\onlinecite{Sie2015},\onlinecite{Sie2017}].
\end{itemize}
 
The large deviation of the parameters $C$, summarized in Tab.~\ref{tab:tab_3}, 
particularly can be explained by i) the sensitivity  
of the OS shifts to the dielectric constant of the medium surrounding the monolayer; 
ii) by different values of the light-matter coupling constants for different TMDs. 
     
\section{Conclusions}
\label{sec:Conclusions}

We have analyzed the OS and BS shifts of 1sA and 1sB excitons 
in WSe$_2$ and MoS$_2$ monolayers induced by ultrashort strong infrared pump 
pulses (FWHM 38~fs, central photon energy $\hbar\omega_\mathrm{pump}=0.62$~eV). 
The observed linear dependence of the shifts with the intensity of the pump pulse
(up to 30 GW/cm$^2$ for WSe$_2$, and up to 50  GW/cm$^2$ for MoS$_2$) has been explained 
in the framework of SBE, based on Dirac-type two-band Hamiltonian with the Coulomb interaction included. 

The theoretical analysis of SBE provided several crucial observations. 
First, we have confirmed the significant importance of the Coulomb interaction 
for correct explanation of the values of the studied shifts. 
Namely, due to the Coulomb effects the shifts are more than twice larger in comparison with 
the results of the simple two-level model considered earlier \cite{Kim2014,Sie2015,Sie2017}. 

Second, the linear dependence of the shifts with the intensity of the pump pulse 
originates from the linear response of the monolayer polarization to the electric field of the pump pulse.
In other words, the studied systems remain in the linear response regime even at high intensities.
This phenomenon can be explained partially by a large bandgap in the system. 
To observe the non-linear effects even larger intensities are needed (more than 50 GW/cm$^2$ in the case 
of MoS$_2$ monolayer).  

Third, our theoretical expressions for the OS and BS shifts contain only the parameters known in the literature.
Therefore additional fitting parameters are not required to evaluate the shifts. 
Theoretical estimates provide fairly good agreement with the experimental results. 
The precision of our theoretical results is limited only by the precision of the parameters, 
such as the Fermi velocities, bandgaps and effective masses of electrons and holes in the system.      

Finally, we have confirmed that the resulting OS and BS shifts occur in different valleys, 
since these effects obey opposite selection rules at the opposite valleys. 
It allows to tune the values of the shifts in each valley separately providing a new tool 
for manipulation of the valley degree of freedom. We have demonstrated that there are 
three parameters that can be used for such a manipulation -- the photon energy of the pump 
pulse $\hbar\omega_\text{pump}$, the intensity of the pump pulse $I_\text{pump}$, and 
the effective dielectric constant $\varepsilon$ of the environment surrounding the monolayer. 
        
\section{Acknowledgments}

We thank B.~Velick\'{y} for his comments to the manuscript.
The authors would like to acknowledge the support by the Czech Science Foundation (project GA18-10486Y) and Charles University (UNCE/SCI/010, SVV-2020-260590, PRIMUS/19/SCI/05).
M. Barto\v{s} acknowledges the support by the ESF under the project CZ.02.2.69/0.0/0.0/20\_079/0017436.

\appendix

\section{Pump power in the monolayers}
\label{app:pump_power}

\begin{figure}[t]
	\centering
	\includegraphics[width=0.9\linewidth]{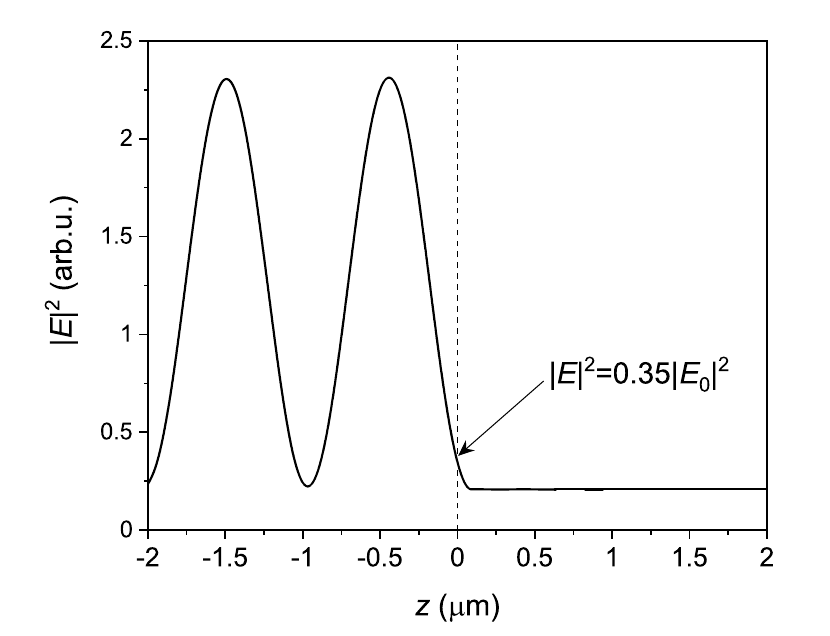}
	\caption{Results of finite-difference time domain simulation showing the normalized distribution of 
	$|\mathbf{E}(z)|^2$  
	at the frequency corresponding to the center pump wavelength of 2 $\mu$m. The power in the monolayer placed at 
	$z=0$ is reduced to 0.35 of the vacuum power of the pump pulse. }
	\label{fig:app_power}
\end{figure}

The presence of the SiO$_2$ layer on the substrate also influences the peak intensity of the pump pulse in the monolayer. To evaluate the peak pump intensity in the monolayer we used the finite-difference time domain (FDTD) simulations, which were performed using a commercial software Lumerical FDTD. The 1D model of the sample consists of four materials: air, monolayer, SiO$_2$ and Si. The monolayer is placed at coordinate $z=0$ and it is simulated as a 1 nm thick layer (much smaller thickness than the wavelength of the driving wave of 2 $\mu$m). 
The dielectric function of the MoS$_2$ monolayer at 0.62 eV is not very well known. We simulated the monolayer using the dielectric function of GaAs, which has a similar band gap. We note that the amplitude of the electric field in the 
1 nm thick layer is almost not influenced by the dielectric function used in the simulation due to the small thickness of the layer.
The SiO$_2$ layer thickness in the simulation is 90 nm, which corresponds to the physical thickness of the oxide layer of our substrates. 
The SiO$_2$ layer is followed by the semiinfinite layer of Si. The dielectric functions of both of these materials are obtained from the material database of the software with values $\varepsilon_{SiO_2}$=2.07 and $\varepsilon_{Si}=11.89$
To cover the small feature of the monolayer in a FDTD simulation, the size of a single mesh element is 
0.25 nm. 
As the output from the FDTD simulations we obtain the time evolution of the electric field amplitude as a function of position.  Using the Fourier transform we evaluate the component at the central frequency corresponding to the photon energy 0.62 eV used in the experiments.
Due to destructive interference between the incident and the reflected wave, the intensity in the monolayer is suppressed to 0.35 times of the vacuum intensity of the pump, which is calculated from the incident pulse and laser beam parameters (see Fig.~\ref{fig:app_power} showing the normalized distribution of the intensity at the frequency corresponding to the center pump wavelength of 2 $\mu$m).

\section{Differential reflectivity of monolayers on SiO$_2$/Si substrate}
\label{app:definition}

We define the differential reflectivity of a monolayer placed on a substrate using the reflectivity measured on the monolayer $R_0(\hbar\omega)$ and the reflectivity of the bare substrate 
$R_\text{sub}(\hbar\omega)$ as $\delta R(\hbar\omega)=(R_0(\hbar\omega)-R_\text{sub}(\hbar\omega))/R_\text{sub}(\hbar\omega)$. The differential reflectivity can be expressed using a simple formula for the case of the monolayer placed on a homogenous substrate as in Ref.~[\onlinecite{McIntyre1971}]:
\begin{equation}
\delta R(\hbar\omega)=-\frac{8\pi d n_1}{\lambda}\text{Im}
\Big\{\frac{\varepsilon_1-\tilde{\varepsilon}_2}{\varepsilon_1-\tilde{\varepsilon}_3}\Big\}
\label{eq:app_reflectivity}
\end{equation}
where $\varepsilon_1$  is the vacuum dielectric constant (we assume that the dielectric constant of air has the same value), $\tilde{\varepsilon}_2=\varepsilon_2'-i\varepsilon_2''$  and  $\tilde{\varepsilon}_3=\varepsilon_3'-i\varepsilon_3''$  are complex dielectric constants of the monolayer and the substrate, $n_1=1$ is the refractive index of air, $d$ is the monolayer thickness and $\lambda$ is the wavelength of the incident light. In the case of nonabsorptive substrate, the complex dielectric constant $\tilde{\varepsilon}_3$  has zero imaginary component and the denominator in the last term of Eq.~(\ref{eq:app_reflectivity}) is real. If there is an electronic resonance (e.g. the exciton state) in the monolayer corresponding to absorption at wavelength $\lambda$, the imaginary part of 
$\tilde{\varepsilon}_2$ is nonzero and negative leading to a positive value of $\delta R(\hbar\omega)$ 
Here we use the same notation as in Ref.~[\onlinecite{McIntyre1971}], where the optical field is given as 
$\mathbf{E}=\mathbf{E}_0\exp(i\omega t-i\mathbf{k\cdot r})$. In the case of absorptive substrate (our case, silicon absorbs light at photon energies of the probe pulse), the sign of differential reflectivity caused by a weakly absorbing monolayer depends on the ratio between dielectric functions of the first and third medium, which for air and silicon gives negative $\delta R(\hbar\omega)$. However, in our experiments, the structure contains also a 90 nm thick layer of SiO$_2$ on the surface of silicon. 
For evaluation of the sign of $\delta R(\hbar\omega)$  in this case we used FDTD simulations using commercial software Lumerical FDTD. The results of these simulations confirm the negative sign of $\delta R(\hbar\omega)$  for our experimental conditions with SiO$_2$/Si substrate. 
This was also confirmed in the differential reflectivity measurements, where we observed the decrease of reflectivity at the resonances corresponding to 1sA and 1sB excitons in WSe$_2$ and MoS$_2$ monolayers 
(see Figs.~\ref{fig:fig_1}(b) and \ref{fig:fig_2}(b)). In the experiment, the reflectivities 
$R_0(\hbar\omega)$ and $\delta R(\hbar\omega)$  are measured by spatially shifting the sample such that the probe beam is incident at the monolayer 
($R_0(\hbar\omega)$) or at the bare substrate ($R_\text{sub}(\hbar\omega)$). The differential reflectivity contains also a broad background, which has been subtracted in Figs.~\ref{fig:fig_1}(b) and \ref{fig:fig_2}(b). 

\begin{figure}[t]
	\centering
	\includegraphics[width=0.9\linewidth]{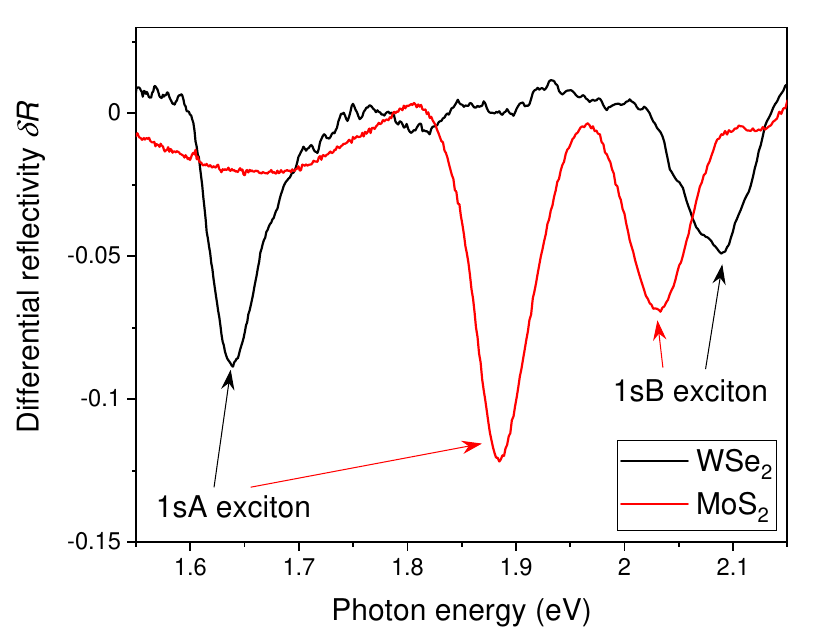}
	\caption{ The experimentally measured differential reflectivities of the samples.}
	\label{fig:app_experimentally}
\end{figure}

The experimentally measured differential reflectivities of the samples used in this study 
are shown in Fig.~\ref{fig:app_experimentally} and confirm the calculation results. T
he resonances corresponding to the excitonic transitions in both materials are visible as dips in the differential reflectivity.

\section{Verification of signal symmetry for different combinations of circular polarizations}
\label{app:verification}

\begin{figure*}[t]
	\centering
	\includegraphics[width=0.8\linewidth]{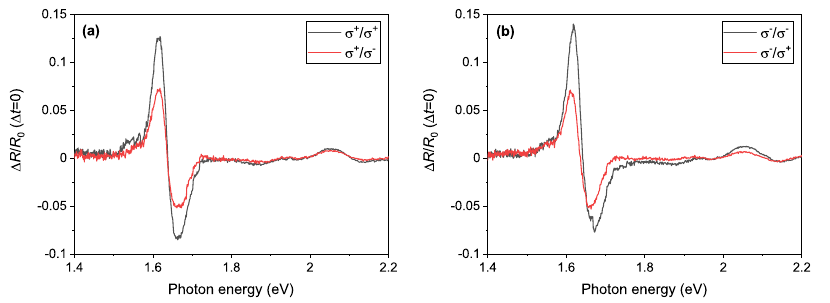}
	\caption{(a) Measured transient reflectivity spectrum of WSe$_2$ monolayer with different combinations of circular polarizations of the pump and probe pulses. Black curves show measurements with co-rotating polarizations 
($\sigma^+/\sigma^+$ and $\sigma^-/\sigma^-$) and red curves show counter-rotating polarizations 
($\sigma^+/\sigma^-$ and $\sigma^-/\sigma^+$).}
	\label{fig:app_verification}
\end{figure*}

In Fig.~\ref{fig:app_verification} we show the transient reflectivity spectra of the WSe$_2$ monolayer in zero time delay between the pump and probe pulses measured with different combinations of circular polarizations of both beams. By this measurement we exclude any experimental artifacts to play role in the observed valley-selective signals. 
This could potentially come from a small displacements of one of the beams when the half wave plate generating the circular polarization is rotated by 
$\pi/2$, which could influence the spatial overlap or the position on the sample. However, because the signals for co-rotating (black curves in Fig.~\ref{fig:app_verification}) and counter-rotating (red curves in 
Fig.~\ref{fig:app_verification}) polarizations are virtually the same for both handednesses of the circular polarization of the pump, such experimental artifacts are excluded.

\section{Derivation of the effective two-band Hamiltonian}
\label{app:two_band_model}

Below we provide the derivation of the Hamiltonian for the bands involved in A exciton transitions.
The case of B excitons is considered analogously.   
For this case, the two-band single-electron Hamiltonian of the monolayer in the $\tau=\pm1$ valley can be written 
up to the quadratic-in-momentum-$\mathbf{k}$ terms as
\begin{align}
H_\mathrm{b}^\tau=&\sum_\mathbf{k}\Big(E_\mathrm{g}+\frac{\Delta_\mathrm{c}}{2}+\gamma_\mathrm{c}k^2\Big)
a^\dag_{\mathbf{k},\mathrm{c},\tau}a_{\mathbf{k},\mathrm{c},\tau}+\nonumber \\+& 
\sum_\mathbf{k}\Big(\frac{\Delta_\mathrm{v}}{2}-\gamma_\mathrm{v} k^2\Big)
a^\dag_{\mathbf{k},\mathrm{v},\tau}a_{\mathbf{k},\mathrm{v},\tau}+\nonumber\\
+&\sum_\mathbf{k}
v(\tau k_x-i k_y)a^\dag_{\mathbf{k},\mathrm{c},\tau}a_{\mathbf{k},\mathrm{v},\tau}
+ \nonumber  \\ +& \sum_\mathbf{k}
v(\tau k_x+i k_y)a^\dag_{\mathbf{k},\mathrm{v},\tau}a_{\mathbf{k},\mathrm{c},\tau}.
\end{align}
Here $a_{\mathbf{k},\mathrm{c},\tau}$ and $a_{\mathbf{k},\mathrm{v},\tau}$ are
the annihilation electron operators in the conduction
and valence bands respectively in the $\tau$ valley with the momentum $\mathbf{k}$ calculated with respect to the 
momentum $\tau\mathbf{K}$, which defines the position of the $\tau$ valley in the Brillouin zone of the TMD monolayer. 
$k$ is the absolute value of the momentum $\mathbf{k}=\mathbf{e}_xk_x+\mathbf{e}_yk_y$. 
For $\tau=1(-1)$ these operators annihilate spin-up(spin-down) electron state in corresponding bands. 
$E_\mathrm{g}$ is the band gap between the conduction and valence bands in the absence of the spin-orbit interaction, 
$\Delta_\mathrm{v}$ and $\Delta_\mathrm{c}$ are the spin-orbit contribution to the energies of the valence and conduction bands respectively. The parameters $\gamma_\mathrm{c}$ and $\gamma_\mathrm{v}$ provide the higher and lower energy bands contribution to the kinetic terms in the considered bands. 
The parameter $v$ defines the interband coupling between the conduction and valence bands 
in the vicinity of K$^\pm$ points. 

We have introduced two-dimensional discrete wave-numbers $\mathbf{k}$, for which we have the following completeness and orthogonality relations  
\begin{align}
\frac{1}{S}&\sum_\mathbf{k} e^{i\mathbf{k}(\mathbf{r}-\mathbf{r}')}=\delta(\mathbf{r}-\mathbf{r}'), \\
\frac{1}{S}&\int_S d^2\mathbf{r}\,e^{i(\mathbf{k}-\mathbf{k}')\mathbf{r}}=\delta_{\mathbf{k}\mathbf{k}'},
\end{align}
where $S$ is the sample's area. The band Hamiltonian can be written as 
$H_\text{b}^\tau=\sum_\mathbf{k}A_\mathbf{k}^{\tau\dag} H_\mathrm{b}^\tau(\mathbf{k})A^\tau_\mathbf{k}$,
where  $A_\mathbf{k}^\tau=
[a_{\mathbf{k},\mathrm{c},\tau},\, a_{\mathbf{k},\mathrm{v},\tau}]^T$ and 
\begin{align}
H_\mathrm{b}^\tau(\mathbf{k})=
 \left[\begin{array}{cccc}
\epsilon_{\mathrm{c},k} & v\tau k e^{-i\tau\phi}\\ 
v\tau ke^{i\tau\phi} & \epsilon_{\mathrm{v},k}\end{array}\right],
\end{align}
where  
$\epsilon_{\mathrm{c},k}=E_\mathrm{g}+\Delta_\mathrm{c}/2+\gamma_\mathrm{c}k^2$, 
$\epsilon_{\mathrm{v},k}=\Delta_\mathrm{v}/2-\gamma_\mathrm{v}k^2$, 
$\phi=\arctan(k_y/k_x)$. The Hamiltonian can be diagonalized by a linear 
transformation  $A_\mathbf{k}^\tau=U^\tau_\mathbf{k}\mathcal{A}_\mathbf{k}^\tau$,
where $\mathcal{A}_\mathbf{k}^\tau=[\alpha_{\mathbf{k},\mathrm{c},\tau}, 
\alpha_{\mathbf{k},\mathrm{v},\tau}]^T$ and 
\begin{align}
U_\mathbf{k}^\tau=
\left[\begin{array}{cc}
\cos(\theta_\mathbf{k}/2) & -\tau e^{-i\tau\phi}\sin(\theta_\mathbf{k}/2) \\ 
\tau e^{i\tau\phi}\sin(\theta_\mathbf{k}/2) & \cos(\theta_\mathbf{k}/2)\end{array}\right]. 
\end{align}
The parameter $\theta_\mathbf{k}$ can be defined from the following equations: 
$\cos\theta_\mathbf{k}=(\epsilon_{\mathrm{c},k}-\epsilon_{\mathrm{v},k})/\epsilon_k$, 
$\sin\theta_\mathbf{k}=vk/\epsilon_k$, with
$\epsilon_k=((\epsilon_{\mathrm{c},k}-\epsilon_{\mathrm{v},k})/2)^2+v^2k^2)^{1/2}$.

Here $\alpha_{\mathbf{k},\mathrm{c},\tau}$ and $\alpha_{\mathbf{k},\mathrm{v},\tau}$ are 
the new fermion annihilation operators, with the same anticommutation relations as 
$a_{\mathbf{k},\mathrm{c},\tau}$ and $a_{\mathbf{k},\mathrm{v},\tau}$.   
After diagonalization the new Hamiltonian reads 
\begin{align}
H_\mathrm{b}^\tau=\sum_\mathbf{k} 
E_{\mathrm{c},k}\alpha^\dag_{\mathbf{k},\mathrm{c},\tau}\alpha_{\mathbf{k},\mathrm{c},\tau} + 
E_{\mathrm{v},k}\alpha^\dag_{\mathbf{k},\mathrm{v},\tau}\alpha_{\mathbf{k},\mathrm{v},\tau}.
\end{align} 
At small $\mathbf{k}$ the band energies take the form 
\begin{align}
E_{\mathrm{c},k}\approx&\,\, 
\epsilon_{\mathrm{c},k}+\frac{v^2k^2}{\bar{E}_\mathrm{g}}=
E_\mathrm{g}+\frac{\Delta_c}{2}+\frac{\hbar^2 k^2}{2m_\mathrm{c}},\\
E_{\mathrm{v},k}\approx&\,\, 
\epsilon_{\mathrm{v},k}-\frac{v^2k^2}{\bar{E}_\mathrm{g}}=
\frac{\Delta_\mathrm{v}}{2}+\frac{\hbar^2 k^2}{2m_\mathrm{v}}
\end{align}
where $m_\mathrm{c}>0$ and $m_\mathrm{v}<0$ are the effective masses of  the carriers in the conduction and valence bands in K$^\pm$ points of monolayer. $E_{\mathrm{c},k}$ and $E_{\mathrm{v},k}$ are the conduction and valence band energies up to the quadratic-in-$\mathbf{k}$ terms. $\bar{E}_g=E_\mathrm{g}+\Delta_\mathrm{c}/2-\Delta_\mathrm{v}/2$ is the single particle band gap of the system. 
 
Since we are interested in exciton effects, we need to add the Coulomb 
interaction terms into our description. 
We will consider only the Coulomb effects which involve the valence and 
conduction bands in each $\tau$ valley separately. 
Then the Coulomb term, responsible for the formation of bright exciton complexes, takes the form 
\begin{align}
H_\mathrm{C}^\tau=&\sum_{\mathbf{k},\mathbf{k}',\mathbf{q}\neq0} \frac{V_\mathbf{q}}{2}\,
a^\dag_{\mathbf{k}+\mathbf{q},\mathrm{c},\tau}a^\dag_{\mathbf{k}'-\mathbf{q},\mathrm{c},\tau}
a_{\mathbf{k}',\mathrm{c},\tau}
a_{\mathbf{k},\mathrm{c},\tau}+\nonumber \\+&
\sum_{\mathbf{k},\mathbf{k}',\mathbf{q}\neq0} \frac{V_\mathbf{q}}{2}\,
a^\dag_{\mathbf{k}+\mathbf{q},\mathrm{v},\tau}a^\dag_{\mathbf{k}'-\mathbf{q},\mathrm{v},\tau}
a_{\mathbf{k}',\mathrm{v},\tau} a_{\mathbf{k},\mathrm{v},\tau}+
\nonumber \\+ &\sum_{\mathbf{k},\mathbf{k}',\mathbf{q}\neq0}V_\mathbf{q}\,
a^\dag_{\mathbf{k}+\mathbf{q},\mathrm{c},\tau}a^\dag_{\mathbf{k}'-\mathbf{q},\mathrm{v},\tau}
a_{\mathbf{k}',\mathrm{v},\tau}a_{\mathbf{k},\mathrm{c},\tau}.
\end{align}
The first and second terms in this expression describe the interactions between electrons in the conduction and valence band, respectively. The last term describes the interaction between electrons in the different bands. 
$V_\mathbf{q}$ is the Fourier transform of the Rytova-Keldysh potential \cite{Cudazzo2011}
\begin{equation}
V_\mathbf{q}=\frac{1}{S}\int d^2\mathbf{r}\, e^{-i\mathbf{qr}}V_{RK}(\mathbf{r})=
\frac{2\pi e^2}{S}\frac{1}{q(qr_0+\varepsilon)},
\end{equation}  
where $r_0$ is the in-plane screening length of the monolayer, and $\varepsilon$ is the average dielectric constant of the medium surrounding the monolayer. Note that $V_\mathbf{q}$ is a function of the absolute value of $\mathbf{q}$, 
i.e. $V_\mathbf{q}=V_q$. Applying the unitary transformation 
$\{a_{\mathbf{k},\mathrm{c},\tau},a_{\mathbf{k},\mathrm{v},\tau}\}\rightarrow \{\alpha_{\mathbf{k},\mathrm{c},\tau},\alpha_{\mathbf{k},\mathrm{v},\tau}\}$, introduced above, one can see that the dominant term of the Coulomb interaction can be obtained by replacing $a_{\mathbf{k},\mathrm{c},\tau}\rightarrow \alpha_{\mathbf{k},\mathrm{c},\tau},a_{\mathbf{k},\mathrm{v},\tau}\rightarrow 
\alpha_{\mathbf{k},\mathrm{v},\tau}$ in $H_\mathrm{C}^\tau$. The other terms contain the small parameter $vk/\bar{E}_\mathrm{g}$ and are omitted  from our consideration.

We present the light-matter interaction as $H_\mathrm{int}^\tau=-\mathbf{P}_\tau\cdot\mathbf{E}$. 
Here $\mathbf{P}_\tau$ is the polarization operator of the system in $\tau$ valley and $\mathbf{E}$ 
is an electric field. The  $j$-th component of the polarization operator in the second quantized form  reads  
\begin{align}
(\mathbf{P}_\tau)_j\approx\frac{-ie\hbar}{m_0\bar{E}_\mathrm{g}} 
\langle\tau\mathbf{K},\mathrm{c}|p_j|\tau\mathbf{K},\mathrm{v}\rangle\sum_\mathbf{k}  
a^\dag_{\mathbf{k},\mathrm{c},\tau}a_{\mathbf{k},\mathrm{v},\tau} + \text{h.c.},
\end{align}
where $e$ and $m_0$ are the bare charge and mass of an electron,
 $|\tau\mathbf{K},\mathrm{v}\rangle$ and $|\tau\mathbf{K},\mathrm{c}\rangle$ are the Bloch states 
of the valence and conduction bands in $\tau$ valley, i.e. in $\tau\mathbf{K}$ point.  
The matrix elements $\langle\tau\mathbf{K}|\widehat{p}_j|\tau\mathbf{K}\rangle$ can be derived in the 
$\mathbf{k\cdot p}$-approximation  
\begin{align}
\langle\tau\mathbf{K},\mathrm{c}|\widehat{p}_x|\tau\mathbf{K},\mathrm{v}\rangle&=\tau v \frac{m_0}{\hbar}, \\
\langle\tau\mathbf{K},\mathrm{c}|\widehat{p}_y|\tau\mathbf{K},\mathrm{v}\rangle&=-iv \frac{m_0}{\hbar}.
\end{align}
Therefore
\begin{align}
\mathbf{P}_\tau=\frac{ev}{i\bar{E}_\mathrm{g}}(\tau\mathbf{e}_x-i\mathbf{e}_y)\sum_\mathbf{k}  
a^\dag_{\mathbf{k},\mathrm{c},\tau}a_{\mathbf{k},\mathrm{v},\tau} + \text{h.c.}.
\end{align}
Hence the light-matter interaction Hamiltonian can be written as
\begin{align}
\label{eq:H_int}
H_\mathrm{int}^\tau=-\mathbf{P}_\tau\cdot\mathbf{E}=\frac{iev}{\bar{E}_\mathrm{g}}(\tau E_x-iE_y)\sum_\mathbf{k}  
a^\dag_{\mathbf{k},\mathrm{c},\tau}a_{\mathbf{k},\mathrm{v},\tau} + \text{h.c.}, 
\end{align}
and for $\sigma^\pm$-polarized light 
$\mathbf{E}=\mathcal{E}(\cos(\omega t)\mathbf{e}_x\pm\sin(\omega t)\mathbf{e}_y)$ one obtains 
\begin{align}
H_\text{int}^\tau=-d^\tau_\mathrm{cv}\mathcal{E}^\tau_\pm(t)\sum_\mathbf{k} 
a^\dag_{\mathbf{k},\mathrm{c},\tau}a_{\mathbf{k},\mathrm{v},\tau}+\text{h.c.},
\end{align}
where $d^\tau_\mathrm{cv}=-iev\tau/\bar{E}_\mathrm{g}=\tau d_\mathrm{cv}$ and $\mathcal{E}^\tau_\pm(t)=\mathcal{E}
e^{\mp i\tau\omega t}$. Note that the structure of the light-matter interaction term has the same form as the corresponding term in the rotating-wave approximation for the two-level problem. However, our expression for 
$H_\mathrm{int}^\tau$ is exact, and has its origin in the helicity-resolved optical selection rules of the monolayer.   
Then, applying the linear transformation 
$\{a_{\mathbf{k},\mathrm{c},\tau},a_{\mathbf{k},\mathrm{v},\tau}\}\rightarrow 
\{\alpha_{\mathbf{k},\mathrm{c},\tau},\alpha_{\mathbf{k},\mathrm{v},\tau}\}$ and introducing the new notation  
$\alpha_{\mathbf{k},\mathrm{c},\tau}\rightarrow \alpha^\tau_\mathbf{k}$, $\beta^\tau_\mathbf{k}\equiv
\alpha^{\tau\dag}_{-\mathbf{k},\mathrm{v},\tau}$, where  
$\beta^\tau_\mathbf{k}$ is the hole annihilation operator in $\tau$ valley. 
The full Hamiltonian takes the form  
\begin{align}
H^\tau=&\sum_\mathbf{k}E_{e,k}\alpha^{\tau\dag}_\mathbf{k}\alpha^\tau_\mathbf{k}+
E_{h,k}\beta^{\tau\dag}_{-\mathbf{k}}\beta^\tau_{-\mathbf{k}}+ \nonumber \\+&
\sum_{\mathbf{k},\mathbf{k}',\mathbf{q}\neq0} \frac{V_\mathbf{q}}{2} (\alpha^{\tau\dag}_{\mathbf{k}+
\mathbf{q}}\alpha^{\tau\dag}_{\mathbf{k}'-\mathbf{q}}\alpha^\tau_{\mathbf{k}'}\alpha^\tau_\mathbf{k}+
\beta^{\tau\dag}_{\mathbf{k}+\mathbf{q}}\beta^{\tau\dag}_{\mathbf{k}'-\mathbf{q}}\beta^\tau_{\mathbf{k}'}\beta_\mathbf{k}^\tau)-\nonumber \\
-&\sum_{\mathbf{k},\mathbf{k}',\mathbf{q}\neq0} V_\mathbf{q} \alpha^{\tau\dag}_{\mathbf{k}+\mathbf{q}}
\beta^{\tau\dag}_{\mathbf{k}'-\mathbf{q}}
\beta^\tau_{\mathbf{k}'}\alpha^\tau_\mathbf{k}-\nonumber \\-& \sum_\mathbf{k} 
(d^\tau_\mathrm{cv}\mathcal{E}^\tau_\pm(t)\alpha^{\tau\dag}_\mathbf{k}\beta^{\tau\dag}_{-\mathbf{k}} + 
\text{h.c.}).
\end{align}
Here, we have introduced the notations 
\begin{equation}
E_{e,k}=E_{\mathrm{c},k}, \quad E_{h,k}=-E_{\mathrm{v},k}+
\sum_{\mathbf{q}\neq 0}V_\mathbf{q}.
\end{equation}
Therefore, the energy of the system containing an electron and a hole with the same momenta 
$\mathbf{k}$ has an energy 
\begin{align}
E_{e,k}+E_{h,k}=\frac{\hbar^2k^2}{2m_e}+
\frac{\hbar^2k^2}{2m_h}+\bar{E}_\mathrm{g}+
\sum_{\mathbf{q}\neq 0}V_\mathbf{q},
\end{align}
where we have introduced the effective electron $m_e\equiv m_\mathrm{c}>0$ and hole $m_h\equiv-m_\mathrm{v}>0$ masses.
The limit  
\begin{align}
\widetilde{E}_\mathrm{g}=\lim_{k\rightarrow 0}(E_{e,k}+E_{h,k})=
\bar{E}_\mathrm{g}+\sum_{\mathbf{q}\neq 0}V_\mathbf{q},
\end{align}
defines the real band gap in the system, renormalized by the Coulomb interaction $V_\mathbf{q}$. 

\section{Interband coupling with photons}
\label{app:quantum_interaction}

The light-matter interaction Hamiltonian (\ref{eq:H_int}) in $\tau=\pm 1$ valley reads 
\begin{equation}
H_\text{int}^\tau=\frac{iev}{\bar{E}_\mathrm{g}}(\tau E_x-iE_y)\sum_\mathbf{k}  
a^\dag_{\mathbf{k},\mathrm{c},\tau}a_{\mathbf{k},\mathrm{v},\tau} + \text{h.c.}
\end{equation} 
In order to have the fully quantized picture of the interband transitions in TMD monolayers
one needs to introduce the second quantized operators of the electric field of the light. 
To this end we consider the second quantized vector potential in the Coulomb gauge 
$\text{div}\mathbf{A}=0$
\begin{equation}
\widehat{\mathbf{A}}(\mathbf{r},t)=\sum_\mathbf{q\alpha} 
\big[\hat{b}_{\mathbf{q}\alpha}\mathbf{A}_{\mathbf{q}\alpha}(\mathbf{r},t)+
\hat{b}^\dag_{\mathbf{q}\alpha}\mathbf{A}^*_{\mathbf{q}\alpha}(\mathbf{r},t)\big],
\end{equation}  
where $\hat{b}_{\mathbf{q}\alpha},\hat{b}^\dag_{\mathbf{q}\alpha}$ 
are the annihilation and creation operators for photons with the wave-vector $\mathbf{q}$ and polarization $\alpha=\pm$.
The vector
\begin{equation}
\mathbf{A}_{\mathbf{q}\alpha}(\mathbf{r},t)=\sqrt{\frac{2\pi\hbar c^2}{L^3\omega_\mathbf{q}}}
\mathbf{e}_\alpha(\mathbf{q}) e^{i(\mathbf{qr}-\omega_\mathbf{q}t)}.
\end{equation}
describes the vector-potential of the $(\mathbf{q},\alpha)$ photon mode.
Here the parameter $c$ is the speed of light, $\hbar$ is the reduced Plank constant and 
$\omega_\mathbf{q}=c|\mathbf{q}|$ is the frequency of a photon with the wave-vector $\mathbf{q}$. 
The polarization vectors are  
\begin{equation}
\mathbf{e}_\pm(\mathbf{q})=\frac{1}{\sqrt{2}}[\mathbf{e}_1(\mathbf{q})\pm i\mathbf{e}_2(\mathbf{q})],
\end{equation}
where $\mathbf{e}_1(\mathbf{q})\perp\mathbf{e}_2(\mathbf{q})$ are real unit vectors perpendicular to $\mathbf{q}$, 
with an additional property $\mathbf{e}_1(\mathbf{q})\times\mathbf{e}_2(\mathbf{q})=\mathbf{q}/|\mathbf{q}|$.
Here ``$\times$'' represents the vector product.  
We supposed that the electromagnetic field is placed in a cubic box with a length $L$, and 
the vector-potential of each mode satisfies the periodic boundary conditions on the opposite walls of the cube.
Therefore the wave-vector $\mathbf{q}$ is parametrized by the set of all integer numbers $(n_x,n_y,n_z)$ as
\begin{equation}
\mathbf{q}=(q_x,q_y,q_z)=\Big(\frac{2\pi n_x}{L},\frac{2\pi n_y}{L},\frac{2\pi n_z}{L}\Big).
\end{equation} 
For this case the operators $\hat{b}_{\mathbf{q}\alpha},\hat{b}^\dag_{\mathbf{k}\beta}$
have the following commutation relations
$[\hat{b}_{\mathbf{q}\alpha},\hat{b}^\dag_{\mathbf{k}\beta}]=\delta_\mathbf{qk}\delta_{\alpha\beta}$, 
$[\hat{b}_{\mathbf{q}\alpha},\hat{b}_{\mathbf{k}\beta}]=
[\hat{b}^\dag_{\mathbf{q}\alpha},\hat{b}^\dag_{\mathbf{k}\beta}]=0$,
where $\delta_{\alpha\beta}$ and 
$\delta_\mathbf{qk}\equiv\delta_{q_x k_x}\delta_{q_y k_y}\delta_{q_z k_z}$ are the Kronecker symbols.

Using these notations the expressions for the operators of the electric  
and magnetic field, given in cgs units, are
\begin{align}
\widehat{\mathbf{E}}=&\frac{i}{c}\sum_\mathbf{q\alpha}\omega_\mathbf{q} 
\big[\hat{b}_{\mathbf{q}\alpha}\mathbf{A}_{\mathbf{q}\alpha}(\mathbf{r},t)-
\hat{b}^\dag_{\mathbf{q}\alpha}\mathbf{A}^*_{\mathbf{q}\alpha}(\mathbf{r},t)\big],\\
\widehat{\mathbf{B}}=&-i\sum_\mathbf{q\alpha}
\big[\hat{b}_{\mathbf{q}\alpha}\mathbf{A}_{\mathbf{q}\alpha}(\mathbf{r},t)\times\mathbf{q}-
\hat{b}^\dag_{\mathbf{q}\alpha}\mathbf{A}^*_{\mathbf{q}\alpha}(\mathbf{r},t)\times \mathbf{q}\big].
\end{align}
 
In these notations the Hamiltonian of the field is 
\begin{equation}
\widehat{H}=\frac{1}{8\pi}\int_{L^3}\!\text{d}^3\mathbf{r}\, \big(\widehat{\mathbf{E}}^2+\widehat{\mathbf{B}}^2\big)=
\sum_\mathbf{q\alpha} \hbar\omega_\mathbf{q}\Big(\hat{b}^\dag_{\mathbf{q}\alpha}\hat{b}_{\mathbf{q}\alpha}
+\frac12\Big),
\end{equation}
the momentum operator of the field is 
\begin{equation}
\widehat{\mathbf{P}}=\frac{1}{4\pi c}\int_{L^3}\!\text{d}^3\mathbf{r}\,\widehat{\mathbf{E}}
\times\widehat{\mathbf{B}}=
\sum_\mathbf{q\alpha} \hbar\mathbf{q}\, \hat{b}^\dag_{\mathbf{q}\alpha}\hat{b}_{\mathbf{q}\alpha},
\end{equation}
and the operator of the spin part of the angular momentum operator of the field is \cite{Messiah1962}  
 \begin{equation}
\widehat{\mathbf{S}}=\frac{1}{4\pi c}\int_{L^3}\!\text{d}^3\mathbf{r}\,\widehat{\mathbf{E}}
\times\widehat{\mathbf{A}}=
\sum_\mathbf{q\alpha} \hbar\frac{\mathbf{q}}{|\mathbf{q}|}\Big(\hat{b}^\dag_{\mathbf{q}+}\hat{b}_{\mathbf{q}+}-
\hat{b}^\dag_{\mathbf{q}-}\hat{b}_{\mathbf{q}-}\Big).
\end{equation}
From the expressions for $\widehat{H}$, $\widehat{\mathbf{P}}$ and $\widehat{\mathbf{S}}$ one concludes  
that the single-photon $(\mathbf{q},\alpha)$-state, created by the operator $\hat{b}^\dag_{\mathbf{q}\alpha}$, 
carries the energy $\hbar\omega_\mathbf{q}$, the momentum $\hbar\mathbf{q}$, 
and the spin angular momentum $\alpha\hbar$. This allows us to interpret the processes 
of absorption and emission of photons in TMD monolayer presented below.

We consider for clarity the monolayer in the $xy$ plane and the case of normal incident a 
light $\mathbf{q}\parallel \mathbf{e}_z$. Then $\mathbf{e}_1(\mathbf{q})=\mathbf{e}_x$, 
$\mathbf{e}_2(\mathbf{q})=\mathbf{e}_y$ and the light-matter interaction term  
takes on the following form   
\begin{widetext}
\begin{equation}
H_\text{int}^\tau=\frac{ev}{\bar{E}_\mathrm{g}}\sqrt{\frac{2\pi\hbar}
{L^3}}\sum_{\mathbf{q}\alpha}\sqrt{\omega_\mathbf{q}} 
\Big[-\Big(\frac{\tau+\alpha}{2}\Big)\hat{b}_{\mathbf{q}\alpha}e^{i\mathbf{qr}-\omega_\mathbf{q}t}+
\Big(\frac{\tau-\alpha}{2}\Big)\hat{b}^\dag_{\mathbf{q}\alpha}e^{-i\mathbf{qr}+\omega_\mathbf{q}t}\Big]
\sum_\mathbf{k}  
a^\dag_{\mathbf{k},\mathrm{c},\tau}a_{\mathbf{k},\mathrm{v},\tau} + \text{h.c.}
\end{equation} 
\end{widetext}
Let us consider the transitions in $\tau=+1$ valley for brevity. 
In this case the Hamiltonian reads  
\begin{widetext}
\begin{equation}
H_\text{int}^{+1}=\frac{ev}{\bar{E}_\mathrm{g}}\sqrt{\frac{2\pi\hbar}
{L^3}}\sum_{\mathbf{q}}\sqrt{\omega_\mathbf{q}} 
\Big[-\hat{b}_{\mathbf{q}+}e^{i\mathbf{qr}-i\omega_\mathbf{q}t}+
\hat{b}^\dag_{\mathbf{q}-}e^{-i\mathbf{qr}+i\omega_\mathbf{q}t}\Big]
\sum_\mathbf{k}  
a^\dag_{\mathbf{k},\mathrm{c},+1}a_{\mathbf{k},\mathrm{v},+1} + \text{h.c.}
\end{equation} 
\end{widetext} 
One can see that the process of creation of the electron-hole pair contains two terms. 
The first term describes the absorption of the photon with the circular polarization $\alpha=+$ and 
energy $\hbar\omega_\mathbf{q}$. 
In this case the angular momentum and energy of the photon are transferred to the crystal causing 
the transfer of an electron from the valence to the conduction band. 
This term defines the well known selection rules for the optical transitions in $\text{K}^+$ point. 

The second term describes the emission of the photon with the circular polarization $\alpha=-$ 
and the energy $\hbar\omega_\mathbf{q}$ with the simultaneous generation of the electron-hole pair. 
Despite conserving the angular momentum, this process  needs an additional external energy. 
Therefore, this process is forbidden for real (on-shell) optical transitions.  
However, this term can play an important role for virtual (off-shell)  
processes in TMDC crystals.

In particular, the latter process becomes relevant for the case 
of high intensities of the incoming light with a large concentration of photons 
in the light beam $n_{\mathbf{q}\alpha}=\langle b_{\mathbf{q}\alpha}^\dag b_{\mathbf{q}\alpha}\rangle/L^3
=N_{\mathbf{q}\alpha}/L^3=\text{const}$.
Taking into account the commutation relations 
\begin{equation}
\frac{\langle b_{\mathbf{q}\alpha} b_{\mathbf{q}\alpha}^\dag\rangle}{L^3}-
\frac{\langle b_{\mathbf{q}\alpha}^\dag b_{\mathbf{q}\alpha}\rangle}{L^3}=\frac{1}{L^3}
\end{equation}
one concludes that  in the limit $L\rightarrow \infty$,  $\langle b_{\mathbf{q}\alpha} b_{\mathbf{q}\alpha}^\dag\rangle\approx 
\langle b_{\mathbf{q}\alpha}^\dag b_{\mathbf{q}\alpha}\rangle$. Then the creation and annihilation operators 
can be considered as commuting objects and can be replaced by complex numbers 
$b_{\mathbf{q}\alpha}\rightarrow e^{i\phi_{\mathbf{q}\alpha}}\sqrt{N_{\mathbf{q}\alpha}}$, 
$b_{\mathbf{q}\alpha}^\dag\rightarrow e^{-i\phi_{\mathbf{q}\alpha}}\sqrt{N_{\mathbf{q}\alpha}}$.
In this limit, the interaction term transforms into   
\begin{widetext}
\begin{equation}
H_\text{int}^{+1}\rightarrow \frac{ev}{\bar{E}_\mathrm{g}}\sqrt{2\pi\hbar}\sum_{\mathbf{q}}\sqrt{\omega_\mathbf{q}} 
\Big[-\sqrt{n_{\mathbf{q}+}}e^{i\mathbf{qr}-i\omega_\mathbf{q}t+i\phi_{\mathbf{q}\alpha}}+
\sqrt{n_{\mathbf{q}-}}e^{-i\mathbf{qr}+i\omega_\mathbf{q}t-i\phi_{\mathbf{q}\alpha}}\Big]
\sum_\mathbf{k}  
a^\dag_{\mathbf{k},\mathrm{c},+1}a_{\mathbf{k},\mathrm{v},+1} + \text{h.c.}
\end{equation} 
\end{widetext}
In this (classical) limit the light of both circular polarizations
becomes coupled with the bands in the $\text{K}^+$ point.  

\section{Polarization of the monolayer by circularly polarized pump light}
\label{app:polarization}

We calculate the polarization $P^\tau_\mathbf{k}$ in $\tau=\pm1$  valleys of the TMD monolayer, induced 
by the $\sigma^+$ polarized pump field. Let us consider $\tau=1$ valley first. 
Eq.~(\ref{eq:bloch_p}) for this case reads
\begin{align}
i\hbar \frac{\partial P_\mathbf{k}}{\partial t}=\hbar e_kP_\mathbf{k}+(2n_\mathbf{k}-1)
\hbar \omega_{R,\mathbf{k}}.
\end{align}
Supposing $n_\mathbf{k}\approx |P_\mathbf{k}^2|\ll 1$ we simplify this equation using the 
expressions for $\hbar e_k$ and 
$\hbar \omega_{R,\mathbf{k}}$ (see Eqs.~(\ref{eq:Rabi_frequency}) and (\ref{eq:energy_parameter}), respectively ) 
\begin{align}
\hbar e_k P_\mathbf{k}&\approx \Big(\widetilde{E}_\text{g}+\frac{\hbar^2k^2}{2m}\Big)P_\mathbf{k}, \\
(2n_\mathbf{k}-1)\hbar \omega_{R,\mathbf{k}}&\approx -d_\text{cv}\mathcal{E}_\text{p}e^{-i\omega_\text{p}t}-
\sum_{\mathbf{q}\neq\mathbf{k}}V_{\mathbf{k}-\mathbf{q}}P_\mathbf{q}.
\end{align}
The substitution $P_\mathbf{k}=p_\mathbf{k}e^{-i\omega_\text{p}t}$ with a time independent $p_\mathbf{k}$ provides 
\begin{equation}
\sum_{\mathbf{k}'}\Big[\Big(\widetilde{E}_\mathrm{g}-\hbar\omega_\text{p}+\frac{\hbar^2k^2}{2m}\Big)
\delta_{\mathbf{k}\mathbf{k}'}-V_{\mathbf{k}-\mathbf{k}'}\Big]p_{\mathbf{k}'}=
\mathcal{E}_\text{p}d_\mathrm{cv}
\end{equation}
We are looking for a solution in the form $p_{\mathbf{k}'}=\sum_{\lambda'} c_{\lambda'}\psi_{\lambda',\mathbf{k}'}$, 
where $\psi_{\lambda',\mathbf{k}'}$ are the eigenfunctions of the equation
\begin{equation}
\sum_{\mathbf{k}'}\Big[\Big(\widetilde{E}_\mathrm{g}+\frac{\hbar^2k^2}{2m}\Big)
\delta_{\mathbf{k}\mathbf{k}'}-V_{\mathbf{k}-\mathbf{k}'}\Big]\psi_{\lambda,\mathbf{k}'}=
\hbar\omega_\lambda\psi_{\lambda,\mathbf{k}'}.
\end{equation}
Then the equation for $p_\mathbf{k}$ transforms into 
\begin{equation}
\sum_{\lambda'}c_{\lambda'} (\hbar\omega_{\lambda'}-\hbar\omega_\text{p})\psi_{\lambda',\mathbf{k}}=
\mathcal{E}_\text{p}d_\mathrm{cv}.
\end{equation}
Multiplying both parts of the equation by $\psi_{\lambda,\mathbf{k}}^*$ and taking the sum over $\mathbf{k}$ 
we obtain
\begin{equation}
c_{\lambda}=\frac{\mathcal{E}_\text{p}d_\mathrm{cv}}{(\hbar\omega_\lambda-\hbar\omega_\text{p})}\sum_\mathbf{k}
\psi_{\lambda,\mathbf{k}}^*=
\frac{\mathcal{E}_\text{p}d_\mathrm{cv}\sqrt{S}}{(\hbar\omega_\lambda-\hbar\omega_\text{p})}
\psi^*_\lambda(\mathbf{r}=0),
\end{equation} 
where we have used the connection between coordinate wavefunction $\psi_\lambda(\mathbf{r})$ normalized as
\begin{equation}
\int_S d^2\mathbf{r}\,\psi^*_\lambda(\mathbf{r})\psi_{\lambda'}(\mathbf{r})=\delta_{\lambda\lambda'}
\end{equation}
and momentum-dependent exciton wave-functions $\psi_{\lambda,\mathbf{k}}$
\begin{equation}
\label{eq:Fourier}
\psi_{\lambda}(\mathbf{r})=\frac{1}{\sqrt{S}}\sum_\mathbf{k} \psi_{\lambda,\mathbf{k}} e^{i\mathbf{kr}}.
\end{equation}
Therefore the value of $p_\mathbf{k}$ takes the form 
\begin{equation}
p_\mathbf{k}=\mathcal{E}_\text{p}d_\mathrm{cv}\sqrt{S}\sum_\lambda 
\frac{\psi_{\lambda,\mathbf{k}}\psi^*_\lambda(\mathbf{r}=0)}{\hbar\omega_\lambda-\hbar\omega_\text{p}}.
\end{equation}
Since we are interested in the lowest energy $1s$-exciton transitions, we can approximate the latter result as 
\begin{equation}
p_\mathbf{k}\approx\mathcal{E}_\text{p}d_\mathrm{cv}\sqrt{S}
\frac{\psi_{1s,\mathbf{k}}\psi_{1s}(\mathbf{r}=0)}{E_{1s}-\hbar\omega_\text{p}}, 
\end{equation}  
where we take into account that $1s$ exciton wave function is real and $\hbar\omega_{1s}=E_{1s}$.
The polarization for the case $\tau=-1$ can be obtained by replacing 
$d_\text{cv}\rightarrow -d_\text{cv}$ and $\omega_\text{p}\rightarrow -\omega_\text{p}$ in the result for 
$\tau=1$ case
\begin{equation}
p_\mathbf{k}=-\mathcal{E}_\text{p}d_\mathrm{cv}\sqrt{S}\sum_\lambda 
\frac{\psi_{\lambda,\mathbf{k}}\psi^*_\lambda(\mathbf{r}=0)}{\hbar\omega_\lambda+\hbar\omega_\text{p}},
\end{equation}
\begin{equation}
p_\mathbf{k}\approx-\mathcal{E}_\text{p}d_\mathrm{cv}\sqrt{S}
\frac{\psi_{1s,\mathbf{k}}\psi_{1s}(\mathbf{r}=0)}{E_{1s}+\hbar\omega_\text{p}}. 
\end{equation}

\section{Estimate of the energy shift for the two-level model}
\label{app:estimation_two_band}

For completeness of the study we derive the OS and BS shifts of the two-level model by considering the limit 
$m\rightarrow\infty$, $me^4\rightarrow 0$ (which nullify the relative kinetic and the Coulomb binding energies of the electron-hole pair, respectively) of the corresponding SBE.
 
We focus on the case $\sigma_\text{p}^+/\sigma_\text{t}^+$, $\tau=1$ and 
$\sigma_\text{p}^+/\sigma_\text{t}^-$ $\tau=-1$ to estimate the OS and BS shift, respectively. 
In the studied limit the equations of motion (\ref{eq:eff_eq_p_plus}) and (\ref{eq:eff_eq_p_minus}) 
for $\tau=\pm 1$ (derived in Secs.~\ref{subsec:tau_plus} and D )
transform into
\begin{equation}
\label{eq:two_level}
(E_0 + 2\tau p^{\tau*}_\mathbf{k}d_\text{cv}\mathcal{E}_\text{p}-\hbar\omega_\text{t})a^\tau_\mathbf{k}
\approx \tau d_\text{cv}\mathcal{E}_\text{t}.
\end{equation}
To derive this equations we have used the corresponding limits of Eqs.~(\ref{eq:h_0}), (\ref{eq:hkk_plus}),
(\ref{eq:hkk_minus}) and replaced $\widetilde{E}_\text{g}\rightarrow E_0$, where $E_0$ is the energy distance
between the levels in the two-level model. $p_\mathbf{k}^\tau$ is the polarization induced by $\sigma^+$ pump pulse in the studied limit. 
Using the results of Appendix~\ref{app:polarization} one obtains 
\begin{equation}
(E_0-\tau\hbar\omega_\text{p})p_\mathbf{k}^\tau=\tau d_\text{cv}\mathcal{E}_\text{p}. 
\end{equation} 
Substitution it into Eq.~(\ref{eq:two_level}) one gets the following result 
\begin{equation}
a^\tau_\mathbf{k}\approx \frac{\tau d_\text{cv}\mathcal{E}_\text{t}}
{E_0+\frac{2|d_\text{cv}|^2\mathcal{E}^2_\text{p}}{E_0-\tau\hbar\omega_\text{p}}-\hbar\omega_\text{t}}.
\end{equation}
Following the analysis from Secs.~\ref{subsec:tau_plus} and \ref{subsec:tau_minus}) 
one obtains the OS 
\begin{equation}
\Delta E_\text{OS}=\frac{2|d_\text{cv}|^2\mathcal{E}^2_\text{p}}{E_0-\hbar\omega_\text{p}}
\end{equation}
and BS shifts
\begin{equation}
\Delta E_\text{BS}=\frac{2|d_\text{cv}|^2\mathcal{E}^2_\text{p}}{E_0+\hbar\omega_\text{p}}
\end{equation}
in the two-level model. 
Both values define the energy scale of the OS and BS shifts. 
For brevity in further calculations, we call them Rabi shifts providing them with
 ``$\pm$'' subscripts
\begin{equation}
\Delta E_{\text{R},\pm}=\frac{2|d_\text{cv}|^2\mathcal{E}^2_\text{p}}{E_0\mp\hbar\omega_\text{p}}.
\end{equation}
  
\section{Estimate of the energy shift for 1s excitons in the monolayer}
\label{app:estimation}

In order to estimate the $1s$ exciton energy shift in the K$^\pm$ points in the presence
of $\sigma^+$ circularly polarized light  
\begin{equation}
\Delta E_\pm=\frac{2|d_\mathrm{cv}|^2\mathcal{E}_\text{p}^2}{E_{1s}\mp\hbar\omega_\text{p}}
\Big[\rho_{1s}+\frac{2\eta_{1s}}{E_{1s}\mp\hbar\omega_\text{p}}\Big],
\end{equation}
we need to evaluate the Rabi shift
\begin{equation}
E_{R,\pm}=\frac{2|d_\mathrm{cv}|^2\mathcal{E}_\text{p}^2}{E_{1s}\mp\hbar\omega_\text{p}},
\end{equation}
and the values of the corresponding parameters
\begin{equation}
\rho_{1s}=\sqrt{S}\psi_{1s}(\mathbf{r}=0)\sum_\mathbf{k}\psi_{1s,\mathbf{k}}^3, 
\end{equation}
\begin{equation}
\eta_{1s}=S[\psi_{1s}(\mathbf{r}=0)]^2
\sum_{\mathbf{k},\mathbf{k}'}V_{\mathbf{k}-\mathbf{k}'}\Big[\psi_{1s,\mathbf{k}}^3\psi_{1s,\mathbf{k}'}-
\psi_{1s,\mathbf{k}}^2 \psi_{1s,\mathbf{k}'}^2\Big].
\end{equation}
To evaluate the parameters $\rho_{1s}$ and $\eta_{1s}$ we express $\psi_{1s,\mathbf{k}}$ via $\psi_{1s}(\mathbf{r})$ 
(inverse relation of Eq.~(\ref{eq:Fourier}))
\begin{equation}
\psi_{1s,\mathbf{k}} =\frac{1}{\sqrt{S}}\int_S d^2\mathbf{r}\,\psi_{\lambda}(\mathbf{r})e^{-i\mathbf{kr}}=
\frac{1}{\sqrt{S}}\varphi(\mathbf{k}),
\end{equation}
substitute the latter into the expressions for $\rho_{1s}$ and $\eta_{1s}$ and take the limit $S\rightarrow \infty$
\begin{equation}
\rho_{1s}=\frac{\psi_{1s}(\mathbf{r}=0)}{(2\pi)^2}\int d^2\mathbf{k}\,\varphi^3(\mathbf{k}), 
\end{equation}
\begin{align}
\eta_{1s}=&\frac{e^2}{(2\pi)^3}[\psi_{1s}(\mathbf{r}=0)]^2\times \nonumber \\ \times&
\iint d^2\mathbf{k}\,d^2\mathbf{k}'\frac{\varphi^3(\mathbf{k})\varphi(\mathbf{k}')-
\varphi^2(\mathbf{k})\varphi^2(\mathbf{k}')}{|\mathbf{k}-\mathbf{k}'|(\varepsilon+r_0 |\mathbf{k}-\mathbf{k}'|)}.
\end{align}
In further calculations we use the variational form of the wave-function
\begin{equation}
\psi_{1s}(\mathbf{r})=\frac{\beta}{\sqrt{2\pi}}e^{-\beta r/2},
\end{equation}
which, as it was demonstrated in Ref.~[\onlinecite{Molas2019}], provides a good approximation for the exciton 
wave-function in TMD monolayers. Therefore
\begin{align}
\varphi(\mathbf{k})=&\frac{\beta}{\sqrt{2\pi}}\int_0^{2\pi} d\phi\int_0^\infty rdr e^{-\beta r/2}e^{-ikr\cos\phi}=
\nonumber \\=&
\frac{4\sqrt{2\pi}\beta^2}{\left(\beta^2+4k^2\right)^{3/2}},
\end{align} 
where we took into account the limit $S\rightarrow \infty$. 
Substituting the obtained expressions into the formula for $\rho_{1s}$ one gets 
\begin{equation}
\rho_{1s}=\frac{\psi_{1s}(\mathbf{r}=0)}{(2\pi)^2}\int d^2\mathbf{k}\,\varphi^3(\mathbf{k})=\frac{16}{7}. 
\end{equation}
Note that this result doesn't depend on the value of the variational parameter $\beta$. 

In order to evaluate $\eta_{1s}$ we make the substitution $\mathbf{k}=\beta \mathbf{x}/2$. 
Then 
\begin{equation}
\varphi(\mathbf{k})=
\frac{4\sqrt{2\pi}}{\beta}\frac{1}{\left(1+x^2\right)^{3/2}}=\frac{4\sqrt{2\pi}}{\beta}\widetilde{\varphi}(\mathbf{x}).
\end{equation} 
Using the dimensionless function $\widetilde{\varphi}(\mathbf{x})$ we present the expression for $\eta_{1s}$ as a 
product of a constant and dimensionless integral, which is a function of the parameter 
$a=\beta r_0/2\varepsilon$
\begin{align}
\eta_{1s}=&\frac{8\beta e^2}{\varepsilon\pi^2}
\iint d^2\mathbf{x}\,d^2\mathbf{y}\frac{\widetilde{\varphi}^3(\mathbf{x})\widetilde{\varphi}(\mathbf{y})-
\widetilde{\varphi}^2(\mathbf{x})\widetilde{\varphi}^2(\mathbf{y})}{|\mathbf{x}-\mathbf{y}|(1+ a|\mathbf{x}-\mathbf{y}|)}= \nonumber \\=&\frac{4\beta e^2}{\varepsilon\pi^2}
\iint d^2\mathbf{x}\,d^2\mathbf{y}\frac{\widetilde{\varphi}(\mathbf{x})\widetilde{\varphi}(\mathbf{y})
[\widetilde{\varphi}(\mathbf{x})-\widetilde{\varphi}(\mathbf{y})]^2}{|\mathbf{x}-\mathbf{y}|(1+ 
a|\mathbf{x}-\mathbf{y}|)}.
\end{align}
The second line of the expression demonstrates that the integral is always positive and decays 
with $a\rightarrow \infty$. Therefore, its value for TMD monolayers is always smaller than for 
the pure Coulomb case $a=0$. 

To perform the calculation one needs to evaluate the integral
\begin{equation}
I(a)=\iint d^2\mathbf{x}\,d^2\mathbf{y}\frac{\widetilde{\varphi}(\mathbf{x})\widetilde{\varphi}(\mathbf{y})
[\widetilde{\varphi}(\mathbf{x})-\widetilde{\varphi}(\mathbf{y})]^2}{|\mathbf{x}-\mathbf{y}|
(1+ a|\mathbf{x}-\mathbf{y}|)}.
\end{equation}
The idea of calculation is based on the introduction of the new parameter $\xi(\phi)=\sqrt{x^2+y^2-2xy\cos\phi}$, 
where $x=|\mathbf{x}|$, $y=|\mathbf{y}|$.
This parameter is nothing but the third length of the triangle defined by the vectors $\mathbf{x}$ and $\mathbf{y}$, 
with the angle $\phi$ between these vectors. The area of this triangle is $\Delta=\frac12xy|\sin\phi|$.
Taking into account that $\widetilde{\varphi}(\mathbf{x})=\widetilde{\varphi}(x)$, 
$\widetilde{\varphi}(\mathbf{y})=\widetilde{\varphi}(y)$, $\xi(\phi)=\xi(-\phi)$ and 
$\xi d\xi=xy\sin\phi d\phi=2\Delta d\phi$
we obtain
\begin{widetext}
\begin{align}
I(a)=&2\pi\int_0^\infty xdx \int_0^\infty ydy \,\widetilde{\varphi}(x)\widetilde{\varphi}(y)
[\widetilde{\varphi}(x)-\widetilde{\varphi}(y)]^2\int_0^{\pi} d\phi \frac{2}{\xi
(1+ a\xi)}=\nonumber \\=&
2\pi\int_0^\infty xdx \int_0^\infty ydy \,\widetilde{\varphi}(x)\widetilde{\varphi}(y)
[\widetilde{\varphi}(x)-\widetilde{\varphi}(y)]^2\int_0^\infty d\xi\frac{\Theta(x,y,\xi)}{\Delta
(1+ a\xi)},
\end{align}
\end{widetext}
where the function $\Theta(x,y,\xi)$ is equal to $1$ when the lengths $x$, $y$, and $\xi$ can form a triangle, 
otherwise it is equal to zero.
Then we use formula no. 6.578.9 from \cite{Gradshtein2007} 
\begin{equation}
\frac{\Theta(x,y,\xi)}{\Delta}=2\pi\int_0^\infty dz\,zJ_0(zx)J_0(zy)J_0(z\xi),
\end{equation}
where $J_\nu(x)$ is the $\nu$-th Bessel function of the first kind. 
We get 
\begin{widetext}
\begin{align}
I(a)=&8\pi^2 \int_0^\infty \frac{d\xi}{1+ a\xi}\int_0^\infty dz\,zJ_0(z\xi)
\Big\{\int_0^\infty xdx J_0(zx)\widetilde{\varphi}^3(x)
\int_0^\infty ydy \,J_0(zy)\widetilde{\varphi}(y)-
\Big[\int_0^\infty xdx J_0(zx)\widetilde{\varphi}^2(x)\Big]^2\Big\}=\nonumber\\
=&8\pi^2 \int_0^\infty \frac{d\xi}{1+ a\xi}\int_0^\infty dz\,zJ_0(z\xi) 
\Big\{\frac{1}{105}e^{-2z}(z(z(z+6)+15)+15)-\frac{1}{64}z^4 K_2^2(z)\Big\}=\nonumber\\
=&8\pi^2 \int_0^\infty \frac{d\xi}{1+ a\xi} \Big\{\frac{10 \xi^{11}+174 \xi^9+\left(1152-35 \sqrt{\xi^2+4}\right) \xi^7+\left(3584-770 \sqrt{\xi^2+4}\right) \xi^5+1470 \sqrt{\xi^2+4} \xi^3}{70 \xi^5 \left(\xi^2+4\right)^{9/2}}+ \nonumber \\
&\qquad\qquad\qquad\qquad+\frac{1260 \sqrt{\xi^2+4}\xi-1680\left(3\xi^4+4\xi^2+3\right)\mathrm{ArcCsch}\left(\frac{2}{\xi}\right)}{70\xi^5\left(\xi^2+4\right)^{9/2}}\Big\},
\end{align}
\end{widetext}
where we used the definition $\widetilde{\varphi}(x)=1/(1+x^2)^{3/2}$.
Here $K_\nu(x)$ is the $\nu$-th modified Bessel function of the second kind. 
Note that the function in braces in the last integral is well localized in the region of small $\xi$, is positive and 
has a maximum at $\xi\approx 1.24$, which makes this function perfectly suited for the numerical estimates. 

Finally, for $a=0$ the integral is evaluated analytically and 
corresponds to the known result for the Coulomb case \cite{Haug2009} 
\begin{align}
I(0)=\pi^2\Big[1-\frac{315\pi^2}{2^{12}}\Big]\approx 2.378.
\end{align}


%

\end{document}